\documentclass[twocolumn,superscriptaddress]{revtex4}

\usepackage[pdftex]{graphicx}
\usepackage{amssymb,amsfonts,amsmath}
\usepackage{verbatim} 
\usepackage{hyperref}

\renewcommand{\thefigure}{A\arabic{figure}}

\newtheorem{theo}{Theorem}
\newtheorem{lem}{Lemma}

\newcommand{\ER}{Erd\H{o}s-R\'{e}nyi }
\mathchardef\mhyphen="2D
\newcommand{\RBR}[3]{$R(#1)\mhyphen B(#2) \mhyphen R(#3)$}
\newcommand{\cutoff}{\mathcal{C}}

\begin{document}

\title{Suppressing cascades of load in interdependent networks\\{\it PNAS Plus Author Summary}}

\author{Charles D. Brummitt}
\email{cbrummitt@math.ucdavis.edu}
\affiliation{University of California, Davis, CA, 95616}
\author{Raissa M. D'Souza}
\email{raissa@cse.ucdavis.edu}
\affiliation{University of California, Davis, CA, 95616}
\affiliation{The Santa Fe Institute, Santa Fe, NM 87501}
\author{E. A. Leicht}
\email{elizabeth.leicht@wolfson.ox.ac.uk}
\affiliation{CABDyN Complexity Centre, University of Oxford, Oxford, United Kingdom}
\affiliation{Sa\"{i}d Business School, University of Oxford, Oxford, United Kingdom}

\maketitle

Modern society depends on increasingly interdependent systems that are prone to widespread failure. Transportation, communication, power grids and other infrastructures support one another and the world's interconnected economies. Barrages of incidents large and small---downed power lines, grounded aircrafts, natural disasters and the like---cause avalanches of repercussions that cascade within and among these systems~\cite{Little2002}. Although interdependence confers benefits, its effect on the risks of individual systems and on the collection of them remains poorly understood. 

Here we analyze how the interconnectivity (interdependence) between networks affects the sizes of their cascades of load shedding. For networks derived from interdependent power grids, we show that interdependence can have a stable equilibrium. An isolated network suppresses its large cascades by connecting to other networks,  but too many interconnections exacerbate its largest cascades---and those of the whole system.  We develop techniques to estimate this optimal amount of interconnectivity, and we examine how differences among networks' capacity and load affect this equilibrium. Our framework advances the current mathematical tools for analyzing dynamics on interdependent (or modular) networks, and it improves our understanding of systemic risk in coupled networks. 

In the basic process we consider, a system contains many elements that shed load to neighboring elements whenever they reach their capacity. This is captured by the classic sandpile model of Bak-Tang-Wiesenfeld, a paradigm for the power law statistics of cascades in many disciplines, from neuronal avalanches to financial instabilities to electrical blackouts \cite{BTW_SOC}. In a basic formulation on a graph of nodes and edges, each node has a capacity for holding grains of sand (interpreted here as load or stress). Grains of sand are dropped randomly on nodes, and whenever a node has more sand than its capacity, it topples and sheds all its sand to its neighbors, which may in turn have too much sand and topple, and so on. Thus dropping a grain of sand can cause an avalanche (cascade) of topplings. These avalanches, like blackouts in power grids~\cite{DobsonChaos2007}, occur in sizes characterized by a power law: they are often small but occasionally enormous.

The Bak-Tang-Wiesenfeld model was originally formulated on a lattice. Given the relevance of 
networked systems, the dynamics have recently been studied on isolated networks, but not yet on interdependent (or modular) networks. Here we study it on two networks with sparse connections between them. Each network models an infrastructure (or a module of one), and the interconnections between them model their interdependence. We explicitly study networks extracted from two interdependent power grids in the southeastern USA and an idealization of them that is more amenable to mathematical study. In this idealization, each node is connected to a node in the other network with probability $p$ (Fig. 1, inset).

\begin{figure}[htbp]
\begin{center}
\includegraphics[width=8.7cm]{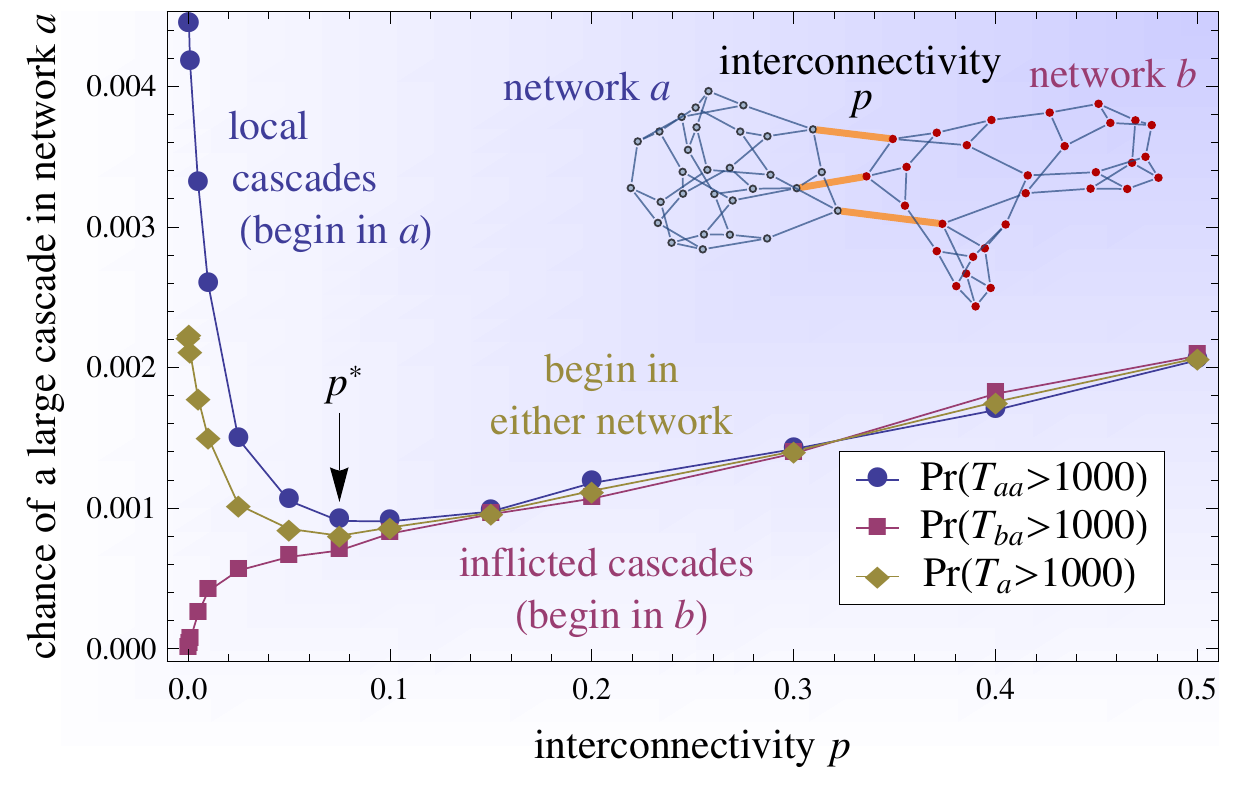}
\caption{The chance that a network $a$ coupled to another network $b$ suffers a cascade larger than half its network (gold curve) has a stable minimum at a critical amount of interconnectivity $p^*$. Networks seeking to mitigate their largest cascades would prefer to build or demolish interconnections to operate at this critical point $p^*$. The blue (red) curve is the chance that a cascade that begins in $a$ ($b$) topples at least 1000 nodes in $a$. Increasing interconnectivity only exacerbates the cascades inflicted from $b$ to $a$ (red), but interestingly it initially suppresses the local cascades in $a$. (From simulations on coupled random 3-regular graphs; the inset depicts a small example with 30 nodes per network and $p=0.1$.)}
\label{locallystabilizing_authorsummary}
\end{center}
\end{figure}

Our main result is that interdependence can have a stable equilibrium (Fig.~\ref{locallystabilizing_authorsummary}). Some interconnectivity is beneficial to an individual network, for the other network acts as a reservoir for extra load. The gold curve of Fig.~\ref{locallystabilizing_authorsummary} shows that the chance of a large cascade in a network can be reduced by $70\%$ by increasing the interconnectivity $p$ from $0.0005$ to $0.075$. Too much interdependence, however, becomes detrimental for two reasons. First, new interconnections open pathways for the neighboring network to inflict additional load. Second, each interconnection augments the system's capacity, making available more load that fuels even larger cascades in each network. As a result, the chance of a large cascade in an individual network eventually increases with interconnectivity $p$, so $p^*$ is a stable minimum.

This second factor above---that new interconnections increase the networks' capacity for load---has global consequences. With more load available, larger cascades in the system as a whole become possible. Therefore networks that interconnect to one another to mitigate their own cascades (Fig.~\ref{locallystabilizing_authorsummary}) may unwittingly cause larger global cascades in the whole system. This is a warning for the interconnections under construction among, for example, different power grids to accommodate long-distance trade and renewable sources of energy~\cite{NPRstory}.

The results in Fig.~\ref{locallystabilizing_authorsummary} show that networks suppressing their largest cascades would seek interconnectivity $p^*$.  However, as shown in the the main article, building interconnections to operate at $p^*$ increases the occurrence of small cascades. Conversely, networks can suppress their smallest cascades the most by seeking isolation, $p=0$. But suppressing their smallest cascades exacerbates their largest ones (left side of Fig.~\ref{locallystabilizing_authorsummary}), just as extinguishing small forest fires can incite large ones and engineering power grids to suppress small blackouts can increase the risk of large ones~\cite{DobsonChaos2007}.

Finally we determine how asymmetry among networks affects the optimal level of interconnectivity that each prefers. For instance, two interconnected power grids may differ in capacity, load, redundancies, demand, susceptibility to line outages, and ages of infrastructure. We capture these differences with a parameter that controls the rates at which cascades begin in either network. We show that in any asymmetric situation the equilibrium will be frustrated, with only one of the grids able to achieve its optimal level of interconnectivity.

Determining how interdependence affects the functioning of networks is critical to understanding the infrastructure so vital to modern society. Whereas others have recently shown that interdependence can lead to alarmingly catastrophic cascades of failed connectivity \cite{Buldyrev2010}, here we show that interdependence also provides benefits, and these benefits can balance the detriments at stable equilibria. We expect that this work will stimulate calculations of critical points in interconnectivity among networks subjected to other dynamics. 
As critical infrastructures such as power grids, transportation, communication, banks and markets become increasingly interdependent, resolving the risks of large cascades and the incentives that shape them becomes ever more important.




\setcounter{section}{0}
\setcounter{figure}{0}

\renewcommand{\thefigure}{\arabic{figure}}

\begin{widetext}
\begin{center}
{\bf \large Suppressing cascades of load in interdependent networks}
\end{center}

\begin{quote}
Understanding how interdependence among systems affects cascading behaviors is increasingly important  across many fields of science and engineering. Inspired by cascades of load shedding in coupled electric grids and other infrastructure, we study the Bak-Tang-Wiesenfeld sandpile model on modular random graphs and on graphs based on actual, interdependent power grids.  Starting from two isolated networks, adding some connectivity between them is beneficial, for it suppresses the largest cascades in each system. Too much interconnectivity, however, becomes detrimental  for two reasons. First, interconnections open pathways for neighboring networks to inflict large cascades. Second, as in real infrastructure, new interconnections increase capacity and total possible load, which fuels even larger cascades. Using a multitype branching process and simulations we show these effects and estimate the optimal level of interconnectivity that balances their tradeoffs. Such equilibria could allow, for example, power grid owners to minimize the largest cascades in their grid.  We also show that asymmetric capacity among interdependent networks affects the optimal connectivity that each prefers and may lead to an arms race for greater capacity. Our multitype branching process framework provides building blocks for better prediction of cascading processes on modular random graphs and on multi-type networks in general.
\end{quote}
\end{widetext}



\begin{figure}[htbp]
\begin{center}
\includegraphics[width=8.7cm]{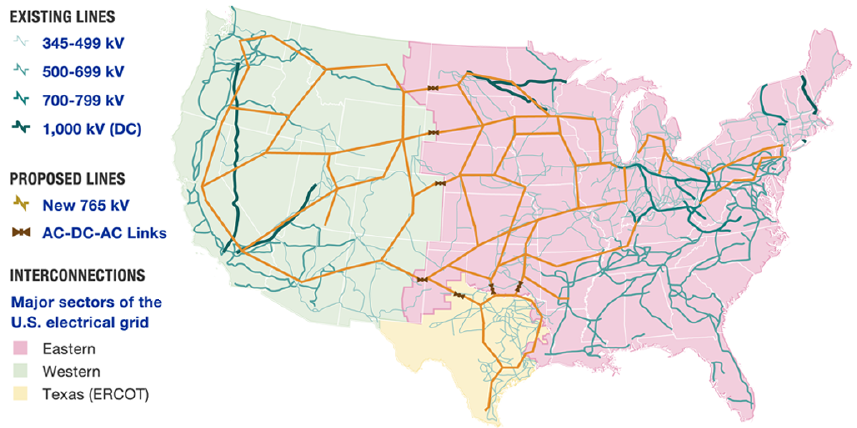}
\caption{The power grid of the continental United States, illustrating the three main regions or ``interconnects''---Western, Eastern and Texas---and new lines (in orange) proposed by American Electric Power to transport wind power. Source: NPR~\cite{NPRstory}.} 
\label{NPRfigure}
\end{center}
\end{figure}

Networks that constitute our critical infrastructure increasingly depend on one another, which enables cascades of load, stress and failures~\cite{Little2002, Rinaldi2004, INLsurvey, Panzieri2008, Amin2000, Grubesic2006}. The water network, for instance, turns turbines and cools nuclear reactors in the electrical grid, which powers the transportation and communications networks that underpin increasingly interdependent global economies. Barrages of disturbances at different scales---volcanic eruptions \cite{volcano_dailymail}, satellite malfunctions \cite{Panzieri2008}, earthquakes, tsunamis, wars \cite{Cederman2003}---trigger cascades of load shedding in interdependent transportation, communication and financial systems. Interdependence can also increase within a particular infrastructure.  The electrical grid of the United States, for example, consists of over 3,200 independent grids with distinct ownership---some private, others public---and unique patterns of connectivity, capacities and redundancies \cite{eia_overview2007}. To accommodate rising demand for electricity, long distance trade of energy~\cite{Lerner:2003}, and new types of power sources~\cite{NPRstory}, interconnections among grids bear ever more load~\cite{aminMRS2008}, and many new high-capacity transmission lines are planned to interconnect grids in the United States and in Europe~\cite{TRJan2011}. Figure \ref{NPRfigure} shows the new interconnections planned to transport wind power~\cite{NPRstory}. Though necessary, these interconnections affect systemic risk in ways not well understood, such as in the power grid, where the modular structure affects its large cascades. For example, the August 14, 2003 blackout, the largest in North American history, spread from a grid in Ohio to one in Michigan, then to grids in Ontario and New York before overwhelming the northeast \cite{Lerner:2003,NERC}.

Researchers have begun to model cascades of load and failure within individual power grids using probabilistic models \cite{DobsonChaos2007}, linearized electric power dynamics \cite{Pepyne2007,HinesChaos2010} and game theory \cite{MotterHICCS2007}. The first models of interdependent grids use simplified topologies and global coupling to find that interconnections affect critical points of cascades~\cite{NewmanHICSS05}, which suggests that they may affect the power law distributions of blackout size~\cite{aminMRS2008,DobsonChaos2007}. Models with interconnections among distinct infrastructure have focused on the spread of topological failures, in which nodes are recursively removed \cite{Buldyrev2010,Parshani2010,Gao2011}, and not on the dynamical processes occurring on these networks. These models find that interdependence causes alarmingly catastrophic cascades of failed connectivity \cite{Buldyrev2010,Parshani2010,Gao2011}. Yet as we show here, interdependence also provides benefits, and these benefits can balance the dangers at stable critical points.  

Here we develop a simple, dynamical model of load shedding on sparsely interconnected networks. 
We study Bak-Tang-Wiesenfeld (BTW) sandpile dynamics \cite{BTW_PRL87,BTW_SOC} on networks derived from real, interdependent power grids and on sparsely coupled, random regular graphs that approximate the real topologies.  Sandpile dynamics are paradigms for the cascades of load, self-organized criticality and power law distributions of event sizes that pervade disciplines, from neuronal avalanches \cite{BeggsPlenz2003,JuanicoJPhysA2007,Ribeiro2010} to cascades among banks \cite{Haldane:2011vo} to earthquakes \cite{Saichev:2004kq}, landslides \cite{Hergarten2003}, forest fires \cite{SinhaRay2000,Malamud1998}, solar flares \cite{Lu1991,Paczuski2005}, and electrical blackouts \cite{DobsonChaos2007}. Sandpile cascades have been extensively studied on isolated networks \cite{sandpile_ER, sandpile_ER_lise,sandpile_wattsstrogatz_1d,sandpile_wattsstrogatz_2d,Goh_sandpile_PRL,Goh_sandpile_PhysicaA,Lee:2005fo}. On interdependent (or modular) networks, more basic dynamical processes have been studied~\cite{vazquez_heterogeneouspopulations, vazquez_structuredpopulations, gleeson_cascademodular, PourboPRE09}, but sandpile dynamics have not.

We use a multitype branching process approximation and simulations to derive at a heuristic level how interdependence affects  cascades of load. Isolated networks can mitigate their largest cascades by building interconnections to other networks, as those networks provide reservoirs to absorb excess load. Build too many interconnections, however, and the largest cascades in an individual network increase in frequency for two reasons: neighboring networks inflict load more easily, and each added interconnection augments the system's overall capacity and load. These stabilizing and destabilizing effects balance at a critical amount of interconnectivity, which we analyze for synthetic networks that approximate interdependent power grids. As a result of the additional load introduced by interconnections, the collection of networks, viewed as one system, suffers larger global cascades---a warning for the increasing interdependence among electrical grids (Fig.~\ref{NPRfigure}), financial sectors and other infrastructure~\cite{NPRstory,aminMRS2008,TRJan2011}. Finally we study the effects of capacity and load imbalance. Networks with larger total capacity inflict larger avalanches on smaller capacity networks, which suggests an arms race for greater capacity. The techniques developed here advance the theoretical machinery for dynamical processes on multi-type networks as well as our heuristic understanding of how interdependence and incentives affect large cascades of load in infrastructure.

\section{Theoretical formulation}
\subsection{Sandpile dynamics}\ 
Introduced by Bak, Tang and Wiesenfeld in 1987 and 1988 \cite{BTW_PRL87,BTW_SOC}, the sandpile model is a well-studied, stylized model of cascades that exhibits self-organized criticality, power laws and universality classes and that has spawned numerous related models with applications in many disciplines (e.g., \cite{Hergarten2003, SinhaRay2000, Lu1991,Paczuski2005,bakBook}). In a basic formulation on an arbitrary graph of nodes and edges, one drops grains of sand (or ``load'') uniformly at random on the nodes, each of which has an innate threshold (or capacity). Whenever the load on a node exceeds its threshold, that node \emph{topples}, meaning that it \emph{sheds} (or moves) its sand to its neighbors. These neighbors may in turn become unstable and topple, which causes some of their neighbors to topple, and so on. In this way, dropping a grain of sand on the network can cause a cascade of load throughout the system---often small but occasionally large. The cascade finishes once no node's load exceeds its capacity, whereupon another grain of sand is dropped, and the process repeats. Probability measures of the size, area, and duration of avalanches typically follow power laws asymptotically in the limit of many avalanches \cite{benhur}.

Many classic versions of the sandpile model~\cite{BTW_PRL87,bakBook,benhur} connect the nodes in a finite, two-dimensional lattice and assign all nodes threshold four, so that a toppled node sheds one sand grain to each of its four neighbors. The lattice has open boundaries, so that sand shed off the boundary  is lost, which prevents inundation of sand.  A few variants of the model on lattices can be solved exactly if the shedding rules have abelian symmetry \cite{benhur}.
 
More recently, sandpile models have been studied on isolated networks, including Erd\H{o}s-R\'{e}nyi graphs \cite{sandpile_ER, sandpile_ER_lise}, scale-free graphs \cite{Goh_sandpile_PRL,Goh_sandpile_PhysicaA,Lee:2005fo}, and graphs generated by the Watts-Strogatz model on one-dimensional \cite{sandpile_wattsstrogatz_1d} and on two-dimensional \cite{sandpile_wattsstrogatz_2d} lattices. A natural choice for the capacities of nodes---which we use here---is their degree, so that toppled nodes shed one grain to each neighbor \cite{Christensen1993,Goh_sandpile_PRL}. Other choices
include identical \cite{Goh_sandpile_PRL}, uniformly distributed from zero to the degree $k$ \cite{Goh_sandpile_PhysicaA}, and $k^{1-\eta}$ for some $0\leq \eta<1$ \cite{Goh_sandpile_PhysicaA,Lee:2005fo}, but all such variants must choose ways to randomly shed to a fraction of neighbors. Shedding one grain to each neighbor is simpler and exhibits the richest behavior~\cite{Goh_sandpile_PRL}. A natural analog of  open boundaries  on finite lattices is to delete grains of sand independently with a small probability $f$ as they are shed. We choose the \emph{dissipation rate of sand} $f$ so that the largest cascades topple almost the entire network.

The mean-field solution of sandpile cascades is characterized by an avalanche size distribution that asymptotically obeys a power law with exponent -$3/2$ and is quite robust to network structure.
(For example, on scale-free random graphs, sandpile cascades deviate from mean-field behavior only if the degree distribution has a sufficiently heavy tail, with power law exponent $2 < \gamma<3$ \cite{Goh_sandpile_PRL}.) Nevertheless, sparse connections among interdependent networks divert and direct sandpile cascades in interesting, relevant ways.

\subsection{Topologies of interacting networks}\  
Here we focus on interdependent power grids and idealized models of them. We obtained topological data on two interdependent power grids---which we label $c$ and $d$---from the US Federal Energy Regulation Commission (FERC) \cite{FERC}. (All data shown here is sanitized to omit sensitive information.) Owned by different utilities but connected to one another in the southeastern USA,\footnote{These two power grids are also connected to other grids, which we ignore in this paper but could model with more types of nodes.} power grids $c$ and $d$ have similar size (439 and 504 buses) but rather different average internal degrees (2.40 and 2.91, respectively). The grids are sparsely interconnected by just eight edges, making the average external node degrees 0.021 and 0.018, respectively. More information on $c,d$ is in Table S1 of the SI. As in other studies, we find that these grids have narrow degree distributions \cite{amaralPNAS2000,albertPRE2004,Wang2010} and are sparsely interconnected to one another~\cite{aminMRS2008}.

To construct idealized versions of the real grids, consider two networks labeled $a$ and $b$. Due to the narrow degree distribution of the real grids, we let network $a$ be a random $z_a$-regular graph (where each node has degree $z_a$) and network $b$ be a random $z_b$-regular graph. These two are then sparsely interconnected as defined below. To define this system of coupled networks more formally, we adopt the multitype network formalism of~\cite{PourboPRE09,LeichtDSouza2009}. Each network $a,b$ has its own degree distribution, $p_a(k_{aa}, k_{ab})$ and $p_b(k_{ba}, k_{bb})$, where, for example, $p_a(k_{aa}, k_{ab})$ is the fraction of nodes in network $a$ with $k_{aa}$ neighbors in $a$ and $k_{ab}$ in $b$. We generate realizations of multitype networks with these degree distributions using a simple generalization of the configuration model: all nodes repeatedly and synchronously draw degree vectors $(k_{oa},k_{ob})$ from their degree distribution $p_o$ (where $o \in \{a,b\}$), until the totals of the internal degrees $k_{aa},k_{bb}$ are both even numbers and the totals of the external degrees $k_{ab}, k_{ba}$ are equal.\footnote{A standard practice in the literature regarding admissible joint degree distributions is that the average inter-degrees $\langle k_{ab} \rangle, \langle k_{ba} \rangle$ of the degree distributions must be equal~ \cite{PourboPRE09,newman_subgraphs}. This is more restrictive than necessary: most any joint degree distributions can be used, as long as conditioning on valid degree sequences (even internal and equal external total degrees) leaves some probability. We derive the \emph{effective} degree distributions, which are convolutions of the input ones, in Lemma \ref{inter_degree_lemma} in the \emph{Appendix}.}

\begin{figure}[htbp]
\begin{center}
\includegraphics[width=8.7cm]{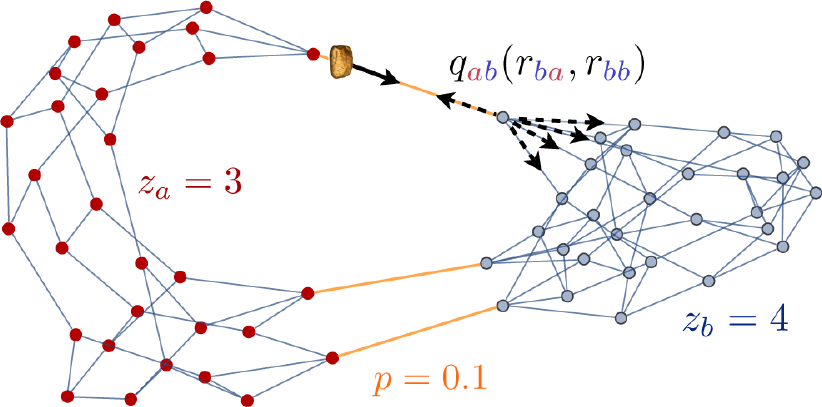}
\caption{Random 3- and 4-regular graphs connected by Bernoulli-distributed coupling with interconnectivity parameter $p=0.1$ (\RBR{3}{0.1}{4}). We also illustrate the shedding branch distribution $q_{ab}(r_{ba,bb})$, the chance that an $ab$-shedding causes $r_{ba}, r_{bb}$ many $ba$-, $bb$-sheddings at the next time step. Note these  random graphs are small and become tree-like when they are large ($\gtrapprox 1000$ nodes).} 
\label{qod_reg3reg4}
\end{center}
\end{figure}

We interconnect the random $z_a,z_b$-regular graphs by \emph{Bernoulli-distributed coupling}: each node receives one external ``edge stub'' with probability $p$ and none with probability $1-p$. Hence the degree distributions are $p_a(z_a, 1)=p, p_a(z_a,0)=1-p$, and $p_b(1,z_b)=p, p_b(0,z_b)=1-p$. We denote this class of interacting networks by the shorthand \RBR{z_a}{p}{z_b}; we illustrate a small example of \RBR{3}{0.1}{4} in Fig.~\ref{qod_reg3reg4}.

\subsection{Measures of avalanche size}\ We are most interested in the \emph{avalanche size distributions} $s_a(t_{a}, t_{b}), s_b(t_{a}, t_{b})$, where, for example, $s_a(t_{a},t_{b})$ is the chance that an avalanche begun in network $a$ (indicated by the subscript on $s_a$) causes $t_{a}, t_{b}$ many topplings in networks $a, b$, respectively. These distributions count the first toppling event, and they are defined asymptotically in that $s_a, s_b$ are frequencies of avalanche sizes after the networks have undergone many cascades. To study $s_a$ and $s_b$, we simulate sandpile avalanches and approximate them using a multitype branching process.

\subsection{Multitype branching process approximation}\ 
In these next two sections, we present an overview of our mathematical formulation, with details left to the {\it Materials and Methods}.   We develop a branching process approximation that elucidates how sandpile cascades spread in interconnected networks, advances theoretical tools for cascades on multitype networks, justifies using this model as an idealization of real infrastructure like power grids, and establishes an open and relevant mathematical challenge. However, readers more interested in the applications of the model may wish to skip to the Results section.

Sandpile cascades on networks can be approximated by a branching process provided that the network is \emph{locally tree-like} (i.e., has few short cycles), so that branches of a nascent cascade grow approximately independently. The interacting networks \RBR{z_a}{p}{z_b} are tree-like provided they are sparse and large enough (with at least several hundred nodes), since the edges are wired uniformly at random. Power grids are approximately tree-like: the clustering coefficient of power grids $c$ and $d$, for example, is $C \approx 0.05$, an order of magnitude larger than an Erd\H{o}s-R\'{e}nyi random graph with equally many nodes and edges, but still quite small. Although tree-based approximations of other dynamical processes work surprisingly well even on \emph{non}-tree-like graphs, power grid topologies were found to be among the most difficult to predict with tree-based theories~\cite{MelnikUnreasonable2011}. Here we find that analytic, tree-based approximations of sandpile dynamics agree well with simulations even on the real power grid topologies (Figs. \ref{matchtheorysimulation_f1}, S7, S8).

Cascades on \emph{interacting} networks require a \emph{multitype} branching process, in which a tree grows according to probability distributions of the number of events of various types generated from seed events. We consider two basic event types, $a$-topplings and $b$-topplings---i.e., toppling events in networks $a$ and in $b$. These simplify the underlying branching process of \emph{sheddings}, or grains of sand shed from one network to another, of which there are four types: $aa$-, $ab$-, $ba$- and $bb$-sheddings. (Note that there is no distinction between topplings and sheddings on one, isolated network, because sand can only be shed from, say, $a$ to $a$.)

A key property of sandpile dynamics on networks, which enables the branching process calculations, is that in simulations the amount of sand on a node is asymptotically \emph{uniformly distributed} from zero to one less its degree (i.e., there is no typical amount of sand on a node)~\cite{Christensen1993,Vespignani1998}. Hence the chance that a grain arriving at a node with degree $k$ topples it equals the chance that the node had $k-1$ grains of sand, which is $1/k$. So sandpile cascades are approximated by what we call \emph{$1/k$-percolation}: the cascade spreads from node $u$ to node $v$ with probability inversely proportional to the degree of $v$. This suggests a direct interpretation for infrastructure:  important nodes have $k$ times more connectivity than unimportant (degree-1) nodes, so they are $k$ times less likely to fail (they are presumably reinforced by engineers). But when important nodes do fail, they cause $k$ times more repercussions (shedded grains of sand). We found some evidence for this in the power flowing through buses (nodes) in power grids: each additional degree correlates with an additional 124 MVA of apparent power flowing through it ($R^2 =0.30$; see Fig. S6 of the SI).

The details of the branching process analysis extend the standard techniques as presented in the \emph{Materials and Methods}. We give here only the crux of the derivation. Suppose a grain of sand is shed from network $o \in \{a,b\}$ to network $d \in \{a,b\}$ (`$o$' for ``origin network'', `$d$' for ``destination network''). What is the chance that this grain shed from $o$ to $d$ (an $od$-shedding) causes $r_{da}$ and $r_{db}$ many grains to be shed from network $d$ to $a$ and from $d$ to $b$, respectively, at the next time step? This probability distribution, denoted $q_{od}(r_{da}, r_{db})$, is the branch (or children) distribution of the branching process for sheddings. Figure \ref{qod_reg3reg4} illustrates $q_{ab}$ as an example. Neglecting degree-degree correlations (the subject of so-called $P(k,k')$ theory~\cite{MelnikUnreasonable2011}), a grain shed from network $o$ to $d$ arrives at an edge stub chosen uniformly at random, so it arrives at a node with degree $p_d(r_{da}, r_{db})$ with probability proportional to $r_{do}$, since that node has $r_{do}$ many edges pointing to network $o$. Using this and the chance of toppling found above ($1/$total degree), we approximate asymptotically that
\begin{align}\label{qod}
q_{od}(r_{da}, r_{db}) &= \frac { r_{do} p_d(r_{da}, r_{db})} {\langle k_{do} \rangle } \frac{1}{r_{da} + r_{db}}
\end{align}
for $r_{da} + r_{db} > 0$, where $\langle k_{do} \rangle$ is the expected number of edges from $d$ to $o$, $\sum_{k_{da}, k_{db}} k_{do} p_d(k_{da},k_{db})$. To normalize $q_{od}$, set
\begin{align}
q_{od}(0,0) &:= 1- \sum_{r_{da} + r_{db} > 0} q_{od}(r_{da}, r_{db}),\label{qab_normalization}
\end{align}
which is the probability that the destination node does not topple (i.e., that it has fewer grains than one less its total degree). 

Note that for an individual, isolated network the analogous branch distribution $q(k)$ simplifies considerably: in the equivalent of Eq.~\eqref{qod} there is a cancelation of $k$ in the numerator of with $1/k$ on the right \cite{Goh_sandpile_PRL, Goh_sandpile_PhysicaA, Lee:2005fo}.  Thus the expected number of children events $\langle q(k) \rangle = \sum_k k \frac{k p(k)}{\langle k \rangle} \frac{1}{k}=1$.  Each seed event gives rise to one child on average, which then gives rise to one child on average, etc., which is called a ``critical" branching process.  (If less than one child on average, the branching process dies out; if more than one it may continue indefinitely.) The branching process approximations of sandpile cascades on the interacting networks studied here---coupled random regular graphs $a,b$ and power grids $c,d$---are also critical, because the principle eigenvalue of the matrix of first moments of the branch distributions is one.

The branching process of sheddings is high dimensional, with four types $aa, ab, ba, bb$ recording origin and destination networks. Transforming the \emph{shedding} branch distributions $q_{od}$ to the \emph{toppling} branch distributions $u_a, u_b$ is easy; the key is that a node topples if and only if it sheds at least one grain of sand (for the details, see the \emph{Materials and Methods}). This also halves the dimensions of the branching process of topplings, simplifying calculations. 

\subsection{Self-consistency equations}\ We analyze implicit equations for the avalanche size distributions $s_a, s_b$ using generating functions \cite{Wilf}. Denote the generating functions associated to the toppling branch distributions $u_a, u_b$ and the avalanche size distributions $s_a, s_b$ by capital letters~$\mathcal{U}$ and $\mathcal{S}$; for example,
\begin{align*}
\mathcal{U}_a(\tau_a, \tau_b) := \sum_{t_a, t_b = 0}^\infty u_a(t_a, t_b) \tau_{a}^{t_a} \tau_{b}^{t_b} \quad \text{for } \tau_{da}, \tau_{db} \in \mathbb{C}.
\end{align*}
The theory of multitype branching processes \cite{Harris1963} implies the \emph{self-consistency equations}
\begin{align}\label{selfconsistencyequations}
\mathcal{S}_a = \tau_a \, \mathcal{U}_a (\mathcal{S}_a, \mathcal{S}_b), \qquad \mathcal{S}_b = \tau_b \, \mathcal{U}_b (\mathcal{S}_a, \mathcal{S}_b),
\end{align}
where each $\mathcal{S}$ is evaluated at $(\tau_a, \tau_b)$. In words, the left-hand equation in \eqref{selfconsistencyequations} says that to obtain the distribution of the sizes of cascades begun in $a$, the cascade begins with an $a$-toppling (hence the $\tau_a$ out front), which causes at the next time step a number of $a$- and $b$-topplings distributed according to $\mathcal{U}_a$, and these topplings in turn cause numbers of $a$- and $b$-topplings distributed according to $\mathcal{S}_a$ and $\mathcal{S}_b$.

\begin{figure}[htbp]
\begin{center}
\includegraphics[width=8.7cm]{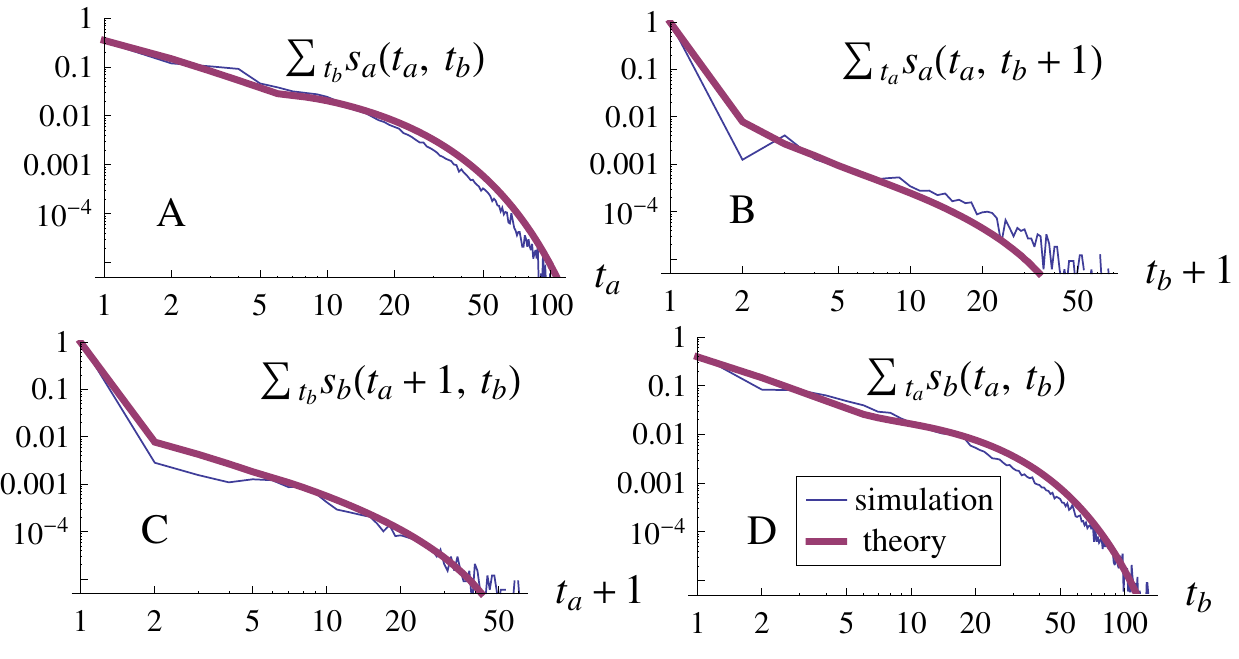}
\caption{The multitype branching process (red curves) approximates simulations of sandpile cascades on the power grids $c,d$ (blue curves) surprisingly well, given that power grids are among the most difficult network topologies on which to predict other dynamics~\cite{MelnikUnreasonable2011}. We plot the four marginalized avalanche size distributions in order to view one-dimensional curves (and here we label power grids $c,d$ as $a,b$, respectively).}
\label{matchtheorysimulation_f1}
\end{center}
\end{figure}

We wish to solve Eqs. \eqref{selfconsistencyequations} for $\mathcal{S}_a$ and $\mathcal{S}_b$, because their coefficients are the avalanche size distributions $s_a, s_b$ of interest. In practice, however, these implicit equations are transcendental and difficult to invert. Instead, we solve Eqs. \eqref{selfconsistencyequations} with computer algebra systems using three methods---iteration, Cauchy integral formula, and multidimensional Lagrange inversion \cite{Good_Lagrange}---to compute exactly hundreds of coefficients; for details, see the \emph{Materials and Methods}. Figure \ref{matchtheorysimulation_f1} shows good agreement between simulation of sandpile cascades on power grids $c,d$ and the branching process approximation (obtained by iterating Eqs. \eqref{selfconsistencyequations} seven times starting from $\mathcal{S}_a=\mathcal{S}_b=1$, with branch distributions calculated from the empirical degree distributions of $c,d$). For more details on the agreement, including how the degrees of nodes with external links account for the characteristic ``blips'' in the avalanche size distributions of the power grids, see Figs. S7 and S8 of the SI. 

These numerical methods are computationally feasible for the probabilities of the smallest avalanches, but we are most interested in the probabilities of the largest avalanches---that is, in the asymptotic behaviors of $s_a(t_a, t_b), s_b(t_a, t_b)$ as $t_a, t_b \rightarrow \infty$. Unfortunately the technique used for an isolated network---an expansion at a singularity of $\mathcal{U}$---fails for sandpile cascades on Bernoulli-coupled random regular graphs and for the power grids, because their generating functions have singularities at infinity and none in the finite plane (see \emph{Materials and Methods}). Generalizing these asymptotic techniques to ``multitype cascades'' with singularities at infinity poses an outstanding mathematical challenge. Nevertheless, three tactics---simulations, computer calculations of coefficients of $\mathcal{S}_a,\mathcal{S}_b$, and analytical calculations of the first moments of the branch and avalanche size distributions---suffice to obtain interesting conclusions about the effect of interdependence on critical cascades of load, as discussed next.

\section{Results}
\subsection{Locally stabilizing effect of interconnections}
\begin{figure}[htbp]
\begin{center}
\centerline{\includegraphics[width=8.7cm]{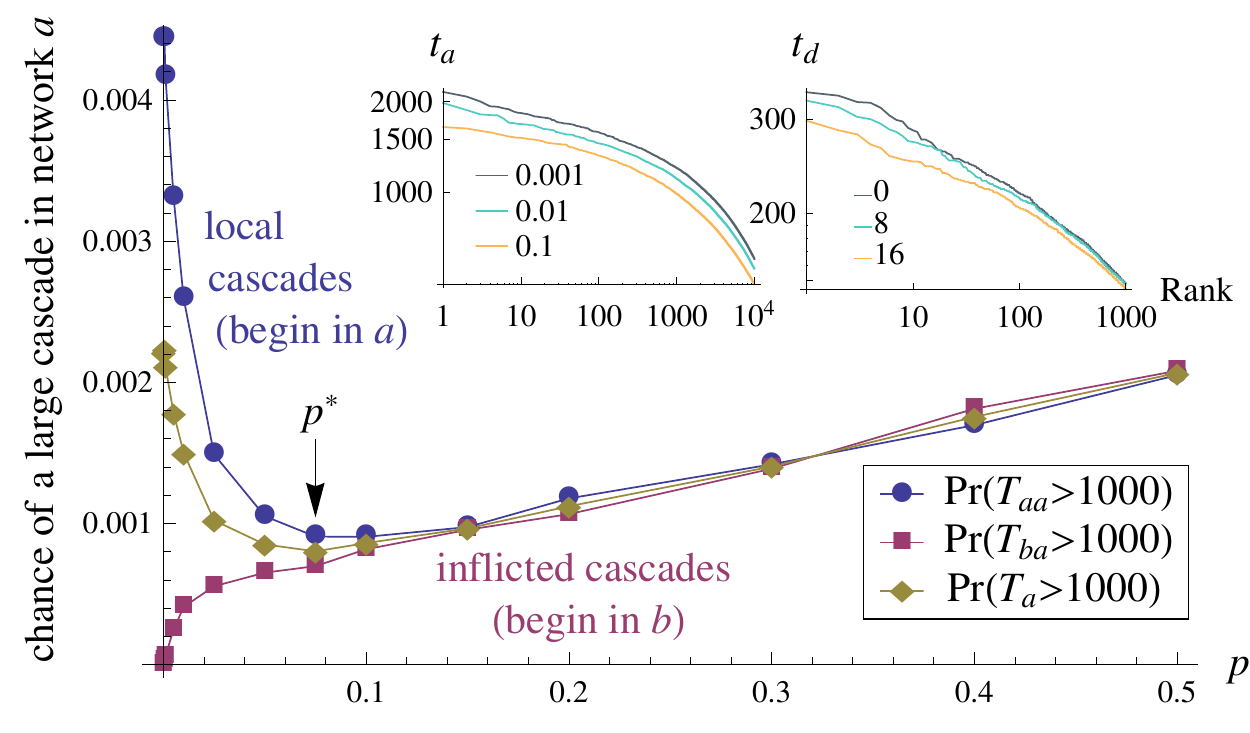}}
\caption{Interconnectivity is locally stabilizing, but only up to a critical point. The main plot, the results of simulations on \RBR{3}{p}{3} ($2\times10^6$ grains, $f=.01$, $2\times10^3$ nodes/network), shows that large local cascades decrease and increase with $p$, while large inflicted cascades only become more likely. Their average (gold curve) has a stable minimum at $p^* \approx 0.075 \pm 0.01$. This curve and its critical point is stable to cutoffs $c$ different from 1000 ($400 \leq x \leq 1500$). Inset figures: rank-size plots on log-log scales of the largest cascades in network $a$ (left) for $p=10^{-3}, 10^{-2}, 10^{-1}$ and in power grid $d$ connected to $c$ by 0, 8 or 16 edges. Both cases are in the regime of sparse interconnectivity (left-hand side of main plot), as greater interconnectivity suppresses large cascades in an individual network.}
\label{locallystabilizing}
\end{center}
\end{figure}

\ 
We first answer the question, would an isolated network suppress its largest cascades of load by connecting to another network? For coupled random regular graphs \RBR{z_a}{p}{z_b}, yes: increasing interconnectivity $p$ suppresses an individual network's largest cascades, but only up to a critical point $p^*$ (Fig.~\ref{locallystabilizing}).

First we introduce notation. For a cascade that begins in network $a$, the random variables $T_{aa}, T_{ab}$ are the sizes of the ``local'' and ``inflicted'' cascades: the number of topplings in $a$ and in $b$, respectively. For example, a cascade that begins in $a$ and that topples 10 $a$-nodes and 5 $b$-nodes corresponds to $T_{aa}=10, T_{ab}=5$. We denote $T_a$ to the be random variable for the size of a cascade in network $a$, without distinguishing where the cascade begins. (We define $T_{ba},T_{bb}, T_b$ analogously.) Dropping sand uniformly at random on two networks of equal size means that avalanches begin with equal probability in either network, so $\Pr(T_a = t_a) = \sum_{t_b}(s_a(t_a, t_b)+s_b(t_a,t_b))/2$. 

In Fig.~\ref{locallystabilizing} we plot the probability of observing a large avalanche in $a$ (that topples at least half of all its nodes) as a function of interconnectivity $p$, as measured in numerical simulations on the \RBR{3}{p}{3} topology. We distinguish between those avalanches that begin in $a$ (blue ``local cascades''), begin in $b$ (red ``inflicted cascades''), or in either network (gold). With increasing interconnectivity $p$, large inflicted cascades from $a$ to $b$ (red curve) increase in frequency due in large part to the greater ease of avalanches traversing the interconnections between networks. More interesting is that increasing interconnectivity suppresses large local cascades (blue curve) for small $p$, but amplifies them for large $p$. The 80\% drop in $\Pr(T_{aa}>1000)$ and 70\% drop in $\Pr(T_a>1000)$ from $p = 0.001$  to $p^* \approx 0.075 \pm 0.01$ are the \emph{locally stabilizing} effects of coupling networks. The left inset to Fig.~\ref{locallystabilizing} is the rank-size plot showing the sizes of the largest avalanches and their decrease with  initially increasing $p$, and the same holds for simulations on the power grids $c$ and $d$ (right inset).\footnote{To adjust interconnectivity between the real power grids $c$ and $d$, we (1) delete the 8 interconnections so that they are isolated, (2) leave the 8 original interconnections, and (3) add 8 additional interconnections in a way that mirrors the empirical degree distribution.} The curve $\Pr(T_a>\cutoff)$ and the location of its critical point $p^*$ in Fig.~\ref{locallystabilizing} is robust to changing the cutoff $\cutoff \in [400,1500]$.  Thus a network such as $a$ seeking to minimize its largest cascades would seek  interconnectivity that minimizes $\Pr(T_a > \cutoff)$, which we estimate to be $p^* \approx 0.075 \pm 0.01$ for \RBR{3}{p}{3} with $2\times10^3$ nodes per network.

This central result appears to be generic: changing the system size, internal degrees, or type of degree distribution (so long as it remains narrow) may slightly change $p^*$ but not the qualitative shape of Fig.~\ref{locallystabilizing} (see SI Figs. S1, S2). Furthermore, this effect of optimal connectivity is unique to interconnected networks: adding edges to a single, isolated network does not reduce the chance of a large cascade (and hence produce a minimum like in Fig.~\ref{locallystabilizing}). (This is expected since $s(t)\sim t^{-3/2}$ for all narrow degree distributions~\cite{Goh_sandpile_PRL}.) 
Whereas adding links within a network can only increase its vulnerability to large cascades, adding links to another network can reduce it. Note that we cannot derive analytical results for Fig.~\ref{locallystabilizing} because the standard techniques for single networks fail for multitype generating functions with singularities at infinity, and inverting Eq.~\eqref{selfconsistencyequations} numerically is practical only for exactly computing the probabilities of small cascades ($T_a <50$) and not large ones ($T_a >10^3$) (see the \emph{Materials and Methods}). These pose open and relevant mathematical challenges for future work.

Intuitively, adding connections between networks diverts load, and that diverted load tends to be absorbed by the neighboring network rather than amplified and returned, as most cascades in isolated networks are small. 
One way to see the diversion of load is in the first moments of the toppling branch distributions $u_a, u_b$. For \RBR{z_a}{p}{z_b}, the average numbers of topplings at the next time step in the same and neighboring networks, respectively, decrease and increase with the interconnectivity $p$ as $\langle u_a \rangle_a = 1-p/(1+z_a), \langle u_a \rangle_b = p / (1+z_b),$
where, for example, $\langle u_a \rangle_a \equiv \sum_{t_a,t_b} t_a u_a(t_a,t_b)$.

However, introducing too many interconnections is detrimental as shown in Fig.~\ref{locallystabilizing}. Interconnections let diverted load more easily return and with catastrophic effect. In addition, each interconnection augments the networks' capacity and hence average load, so that large avalanches increase in frequency in individual networks and in the collection of networks, as discussed next. 

\subsection{Globally destabilizing effect of interconnections}
\begin{figure}[htbp]
\begin{center}
\centerline{\includegraphics[width=8.7cm]{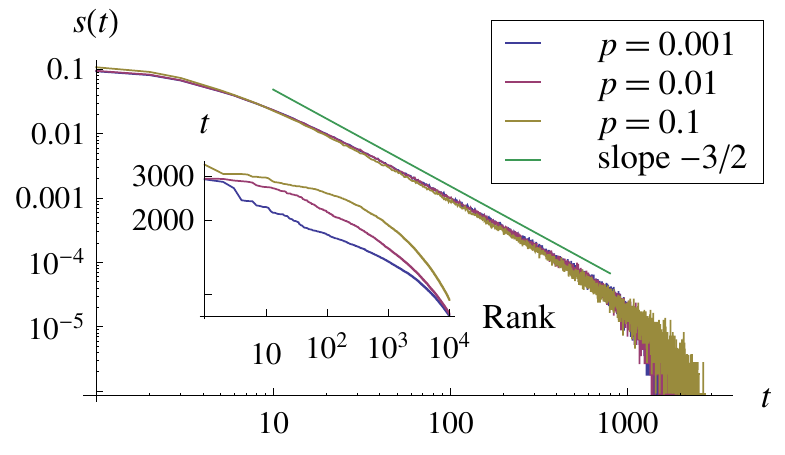}}
\caption{Increasing the interconnectivity $p$ between two random 3-regular graphs extends the tail of the total avalanche size distribution $s(t)$, which does not distinguish whether nodes are in network $a$ or $b$ but considers them as one network. The inset shows a rank-size plot on log-log scales of the number of topplings $t$ in the largest $10^4$ avalanches (with $2\times10^6$ grains of sand dropped), confirming that adding more edges between random 3-regular graphs enlarges the largest global cascades by an amount on the order of the additional number of interconnections.}
\label{totalavalanchesize}
\end{center}
\end{figure}

\ Adding interconnections amplifies large global cascades. That is, the largest avalanches in the collection of networks---viewed as one system with just one type of node---increase in size with interconnectivity. Here we are interested in the \emph{total avalanche size distribution} $s(t)$, the chance of $t$ topplings overall in a cascade.  Figure~\ref{totalavalanchesize} shows the extension of the right-hand tail of $s(t)$ in simulations on \RBR{3}{p}{3} with increasing interconnectivity $p$. The rank-size plot inset shows more explicitly that the largest avalanches increase with $p$. Similar results occur in simulations on power grids $c,d$ (see page 2 of the SI).
 
What amplifies the global cascades most significantly is the increase in total capacity (and hence average load available for cascades) and not the increased interdependence between the networks. (Recall that capacities of nodes are their degrees, so introducing new edges between networks increments those nodes' degrees and capacities.) To see this effect on coupled random regular graphs, we perform the following rewiring experiment. Beginning with two isolated random regular graphs, each node changes one of its internal edge stubs to be external with probability $p$. The degree distributions become, for example, $p_a(z_a-1,1)=p, p_a(z_a,0)=1-p$, which we call ``Correlated-Bernoulli coupling'' because the internal and external degrees are not independent. Figure S4 of the SI shows that the largest global avalanches are not significantly enlarged with increasing ``rewired interconnectivity'' $p$ for random 3-regular graphs with such coupling. Furthermore, the enlargement of the largest cascades observed in the rank-size plot in the inset of Fig.~\ref{totalavalanchesize} is on the order of the extra average load resulting from the additional interconnectivity (and the same holds for simulations on the power grids). 
 
The amplification of global avalanches, though relatively small, is relevant for infrastructure: additional capacity and demand often accompany---and even motivate---the construction of additional interconnections~\cite{NPRstory,TRJan2011}. Furthermore, in reality it is more common to augment old infrastructure with new interconnections as in Fig.~\ref{NPRfigure} (Bernoulli coupling) rather than to delete an existing internal connection and rewire it to span across networks (Correlated-Bernoulli coupling). Thus, building new interconnections augments the entire system's capacity, and hence average load, and hence largest cascades. 

\subsection{Interconnectivity that mitigates cascades of different sizes}
\begin{figure}[htbp]
\begin{center}
\centerline{\includegraphics[width=8.7cm]{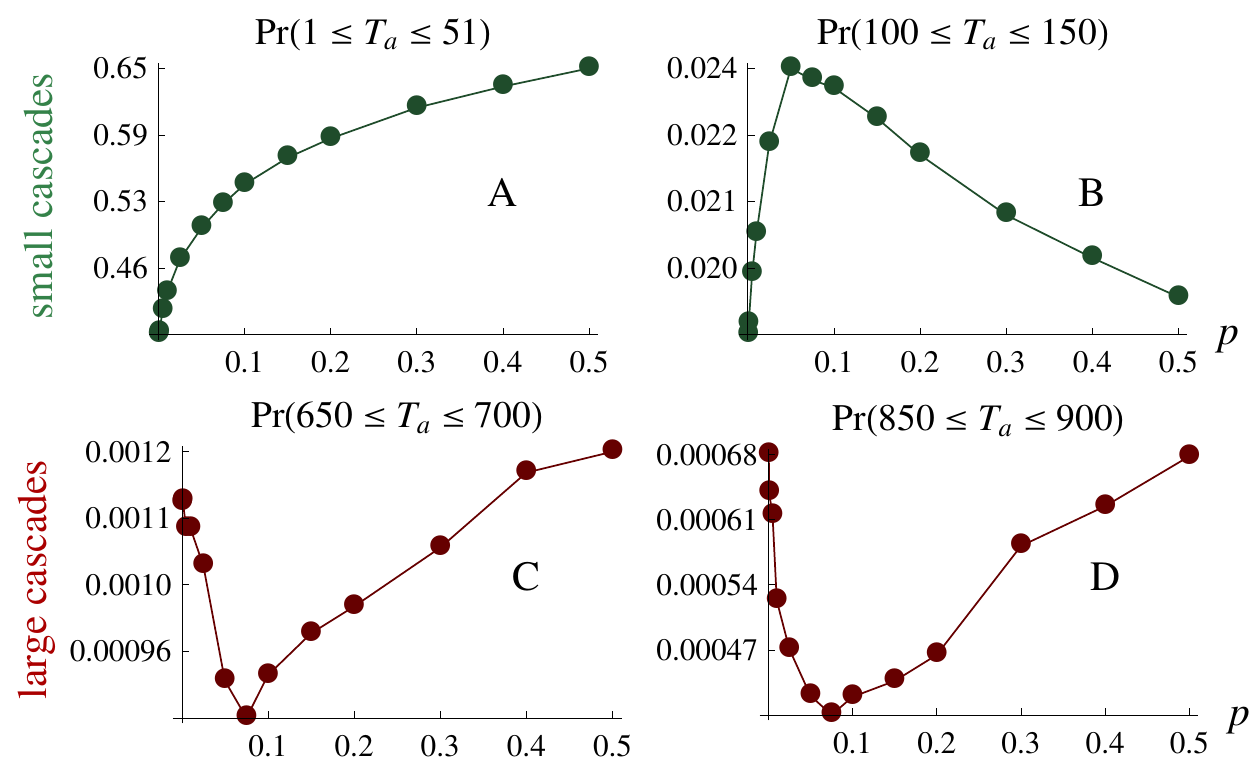}}
\caption{Networks mitigating the smallest cascades $1 \leq T_a \leq 51$ (A) seek isolation $p=0$, while networks suppressing intermediate cascades $100 \leq T_a \leq 150$ (B) seek isolation $p=0$ or strong coupling $p=1$, depending on the initial interconnectivity $p$ in relation to the unstable critical point $p^* \approx 0.05$. But networks like power grids mitigating large cascades (C, D) would seek interconnectivity at the stable equilibrium $p^* \approx 0.075 \pm 0.01$. The bottom figures and the location of $p^*$ are robust to changes in the window $\ell \leq T_a \leq \ell + 50$ for all $400 \leq \ell \leq 1500$.}
\label{concavecurves}
\end{center}
\end{figure}

\ 
Figure \ref{locallystabilizing} shows that networks seeking to mitigate their large avalanches seek optimal interconnectivity $p$. Networks mitigating their small or intermediate cascades would seek different optimal interconnectivity $p$, as shown in Fig.~\ref{concavecurves} (the results of simulations on \RBR{3}{p}{3}). Figure \ref{concavecurves}A shows that the probability of a small cascade in network $a$ ($1 \leq T_a \leq 51$) increases monotonically with $p$, so that networks mitigating the smallest cascades seek isolation, $p=0$. By contrast, the chance of a cascade of intermediate size, $100 \leq T_a \leq 150$ (Fig.~\ref{concavecurves}B), has a local maximum at $p^* \approx 0.05$, so networks mitigating intermediate cascades would demolish all interconnections ($p=0$) or build as many as possible ($p=1$). By contrast, the largest cascades $400 \leq T_a \leq 1500$ (Figs. \ref{concavecurves}C, \ref{concavecurves}D) occur with minimal probability at $p^* \approx 0.075 \pm 0.01$.  For more plots showing the change in concavity and the robustness of the stable critical point $p^*$ for large cascades, see Fig. S3.

Other models of cascades in power grids conclude that upgrading and repairing the system to mitigate the smallest blackouts may increase the risk of the largest blackouts~\cite{DobsonChaos2007}. Similarly, extinguishing small forest fires, a common policy in the 20th century, increases forest vegetation and thus the risk of large forest fires---a phenomenon known as the ``Yellowstone effect'' \cite{Malamud1998}. The results here augment these conclusions to include interconnectivity. Networks building interconnections in order to suppress their large cascades cause larger global cascades (by an amount on the order of the additional capacity). Networks suppressing their small or intermediate cascades may seek isolation ($p=0$), which amplifies their large cascades, or (for intermediate cascades) strong coupling ($p=1$), which amplifies both large local and global cascades.

\subsection{Capacity disparity}\ Two networks that are interdependent are rarely identical, as are the \RBR{3}{p}{3} topologies studied thus far, so we determine the effect of capacity disparity on cascades. As the capacities of nodes are their degrees, we study \RBR{z_a}{p}{z_b} with different internal degrees, $z_a \neq z_b$. We find that interdependence is more catastrophic for smaller capacity networks, in that they suffer comparatively larger inflicted cascades.  They still prefer some interconnectivity, but less than the higher capacity network.

Using the theoretical branching process approximation, we compute how much larger inflicted cascades are from high- to low-capacity networks. Differentiating Eqs.~\eqref{selfconsistencyequations} with respect to $\tau_a, \tau_b$ and setting $\tau_a=\tau_b=1$ yields four linear equations for the first moments of the avalanche size distributions $s_a, s_b$ in terms of the first moments of the branch distributions $u_a, u_b$. For \RBR{z_a}{p}{z_b}, the four first moments of $s_a, s_b$ are all infinite, as expected, because in isolation these networks' avalanche size distributions are power laws with exponent $-3/2$~\cite{Goh_sandpile_PRL,Goh_sandpile_PhysicaA,Lee:2005fo}. Nevertheless, we can compare the rates at which the average inflicted cascade sizes diverge by computing their ratio
\begin{align}\label{inflictedratio}
\frac {\langle s_a \rangle_b} {\langle s_b \rangle_a} = \frac{1+z_a}{1+z_b},
\end{align}
where, e.g., $\langle s_a \rangle_a \equiv \sum_{t_a,t_b} t_a s_a(t_a,t_b)$ (see SI for the derivation). Thus $z_b>z_a$ implies that the inflicted cascades from $b$ to $a$ are larger on average than those from $a$ to $b$.
Fig. S9 of the SI shows qualitative agreement with simulations. 

As a result of Eq.~\eqref{inflictedratio}, the low-capacity network prefers less interconnectivity than the high-capacity network. In simulations of \RBR{3}{p}{4}, for instance, low capacity network $a$ prefers $p_a^* \approx 0.05$, whereas high-capacity $b$ prefers $p_b^* \approx 0.3$. For systems like power grids seeking to mitigate their cascades of load, these results suggest an arms race for greater capacity to fortify against cascades inflicted by neighboring networks. 

\subsection{Incentives and equilibria in power grids}\ 
Since different networks have unique susceptibilities to cascades (due to capacity disparity, for example), equilibria among real networks are more nuanced than on identical random graphs. Next we explore how the level of interconnectivity and load disparity affect sandpile cascades on the power grids $c$ and $d$. (Although sandpile dynamics do not obey Ohm's and Kirchoff's laws nor the flow of load from sources to sinks,  as in physical power flow models (e.g.,~\cite{Pepyne2007,HinesChaos2010}), they do closely resemble some engineers' models of blackouts, and blackout data show evidence of criticality and power laws \cite{DobsonChaos2007}.) To interpret results, we suppose that the owners of the power grids $c,d$ are rational, in that they wish to mitigate their largest cascades but care little about cascades overwhelming neighboring grids.

To capture different amounts of demand, numbers of redundancies, ages of infrastructure,  susceptibility to sagging power lines~\cite{Pepyne2007,HinesChaos2010}, and other factors that affect the rate at which cascades of load shedding and failures begin in each network, we introduce a \emph{load disparity} parameter $r$ as follows. Each node in $c$ is $r$ times more likely than a node in $d$ to receive a new grain of sand. Increasing $r$ intensifies the load on grid $c$, the rate at which cascades begin there, and the sizes of the largest inflicted cascades from $c$ to $d$. The larger $r$ is, the more volatile power grid $c$ becomes.

Given the spatial structure of the power grid networks, there is no principled way to add arbitrarily many interconnections between them. However, three different levels of interconnectivity are natural: (1) delete the eight interconnections so that $c$ and $d$ are isolated, (2) leave the eight original interconnections, and (3) add eight additional interconnections in a way that mirrors the empirical degree distribution. 

\begin{figure}[htbp]
\begin{center}
\includegraphics[width=8.7cm]{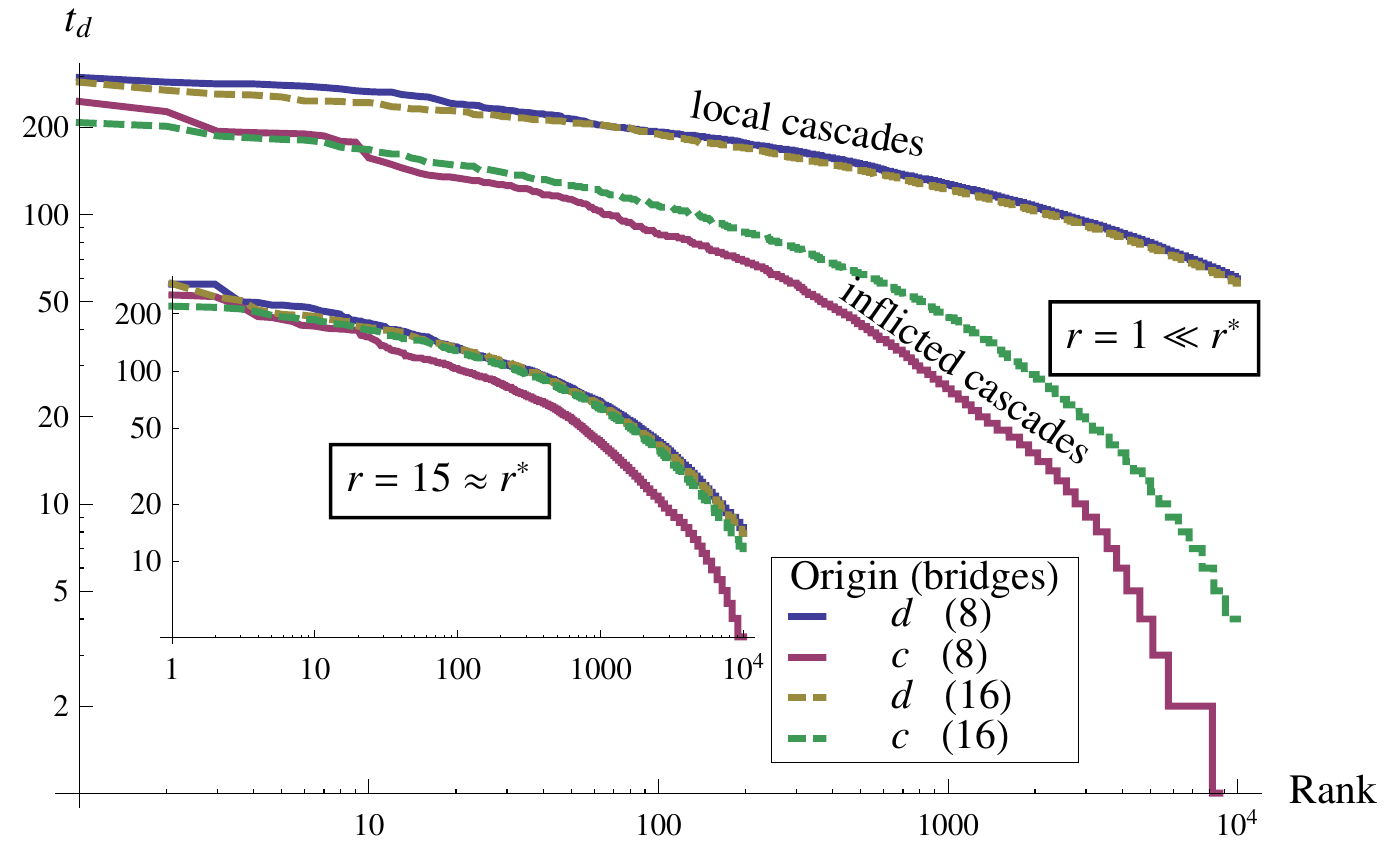}
\caption{The critical load disparity at which inflicted cascades in $d$ become equally large as its local cascades is $r^* \approx 15$ (for 16 interconnections (``bridges'')). (For 8 interconnections, $r^* > 20$; see Fig.~\ref{rankplot_cd_r10andr20}). Here we show a rank-size plot in log-log scales of the largest $10^4$ avalanches in power grid $d$, distinguishing whether they begin in $c$ (``inflicted cascades'') or in $d$ (``local cascades''), for 8 and 16 interconnections (solid, dashed curves), for $r =1$ (main), $r=15$ (inset), in simulations with dissipation of sand $f=0.05$, $10^6$ grains dropped (after $10^5$ transients).} 
\label{rankplot_cd_r1andr15}
\end{center}
\end{figure}

Figure  \ref{rankplot_cd_r1andr15} shows that there are two ways to amplify the largest inflicted cascades in $d$. The first is to increase the number of interconnections (compare red to green-dashed curve). The second is to increase $r$ (compare distance from green to gold curve in the main and in the inset). At a critical $r^*$, the largest inflicted cascades in $d$ that begin in $c$ (red and green curves) are equally large as the largest local cascades in $d$ that begin in $d$ (blue and gold curves). For 16 interconnections, we estimate $r^* \approx 15$ (inset of Fig.~\ref{rankplot_cd_r1andr15}), and the inflicted cascades are larger and smaller, respectively, for $r=10, 20$ (Fig.~S5)---indicating that inflicted cascades begin to dominate local cascades at $r^* \approx 15$. The actual load disparity between power grids $c$ and $d$ is $r \approx 0.7$, which we estimate by computing the average power incident per node in simulation output from FERC \cite{FERC}. (There are, however, interdependent power grids in the southeastern USA with $r > 15$.) Since $r=0.7$, the load is greater on grid $d$, so grid $d$ prefers more interconnections and $c$ prefers fewer than if $r$ were 1. Consequently, any equilibrium between the two grids is \emph{frustrated} (or \emph{semi-stable}): only one grid can achieve its desired interconnectivity. 

\section{Discussion}
We have presented a comprehensive study of sandpile cascades on interacting networks to obtain a deeper understanding of cascades of load on interdependent systems, showing both the benefits and dangers of interdependence. 
We combine a mathematical framework for multitype networks \cite{PourboPRE09,LeichtDSouza2009} with models of sandpiles on isolated networks~\cite{sandpile_ER,sandpile_ER_lise,sandpile_wattsstrogatz_1d,sandpile_wattsstrogatz_2d,Goh_sandpile_PRL,Goh_sandpile_PhysicaA,Lee:2005fo} to derive a multitype branching process approximation of cascades of load between simple interacting networks and between real power grids. We show that some interdependence is beneficial to a network, for it mitigates its largest avalanches by diverting load to neighboring networks. But too much interconnectivity opens new avenues for inflicted load and adds capacity that fuels even larger cascades. The benefits and detriments in mitigating large avalanches balance at a critical amount of interconnectivity, which is a stable equilibrium for coupled networks with similar capacities.  For coupled networks with asymmetric capacity, the equilibria will be frustrated, in that the networks prefer different optimal levels of interconnectivity. 

We also show that tuning interconnectivity to suppress cascades of a certain range of sizes 
amplifies the occurrence of cascades in other ranges. Thus a network mitigating its small  avalanches amplifies its large ones (Fig.~\ref{locallystabilizing}), and networks suppressing their own large avalanches amplify large ones in the whole system (Fig.~\ref{totalavalanchesize}). Similarly it has been found that mitigating small electrical blackouts and small forest fires appears to increase the risk of large ones~\cite{DobsonChaos2007,Malamud1998}. Furthermore, the amplification of global cascades due to  the increase in capacity (Fig.~\ref{totalavalanchesize}) is a warning for new interconnections among power grids (Fig.~\ref{NPRfigure}) \cite{NPRstory, TRJan2011}, financial sectors and other infrastructure.

These findings suggest economic and game-theoretic implications for infrastructure. (Note that here we consider the strategic interactions of \emph{entire networks} rather than of individual nodes in one network, which is more standard, e.g., \cite{MatthewJackson}.) For example, a power grid owner wishing to mitigate the largest cascades in her grid would desire some interconnections to other grids, but not too many. However, what benefits individual grids can harm society: grids building interconnections amplify global cascades in the entire system. More detailed models that combine results like these with economic and physical considerations of electrical grids and with costs of building connections may provide more realistic estimates of optimal levels of interconnectivity. Our framework---which models a dynamical process on stable, underlying network topologies---could also be combined with models of topological failure in interdependent networks~\cite{Buldyrev2010,Parshani2010,Gao2011}. Those studies conclude that systemic risk of connectivity failure increases monotonically with interdependence (``dependency links'').  Whether suppressing cascades of load or of connectivity failures is more important might suggest whether to interconnect some ($p^*>0$) [Fig.~\ref{locallystabilizing}] or none ($p=0$)~\cite{Buldyrev2010,Parshani2010,Gao2011}, respectively. Our results are consistent with recent work showing that an intermediate amount of connectivity minimizes risk of systemic default in credit networks~\cite{Battiston2009}, in contrast to Refs.~\cite{Buldyrev2010,Parshani2010,Gao2011} and the more traditional view that risk increases monotonically with connectivity in credit networks (e.g., \cite{Allen2000}).

This work also advances our mathematical understanding of dynamical processes on multitype networks. Since networks with one type of node and edge are impoverished views of reality, researchers have begun to study dynamical processes on multitype networks, such as on modular graphs \cite{vazquez_heterogeneouspopulations, vazquez_structuredpopulations, gleeson_cascademodular,PourboPRE09}. Here we derive a branching process approximation of sandpile cascades on multitype networks starting from the degree distributions, and we discuss open problems in solving for the asymptotic behavior of the generating functions' coefficients, which elude current techniques for isolated networks. We expect that the computational techniques used here to solve multidimensional generating function equations, such as multidimensional Lagrange inversion \cite{Good_Lagrange}, will find other uses in percolation and cascades in multitype networks.  Finally, in the \emph{Appendix} we derive the effective degree distributions in multitype networks, which expands the admissible degree distributions that others have considered.  
The machinery we develop considers just two interacting networks, $a$ and $b$, or equivalently one network with two types of nodes. However, this extends to finitely many types, which may be useful for distinguishing types of nodes---such as buses, transformers and generators in electrical grids---or for capturing geographic information in a low-dimensional way without storing explicit locations---such as buses in the interiors of power grids and along boundaries between them.

Here we have focused on mitigating large avalanches in modular networks, but other applications may prefer to amplify large cascades, such as the adoption of products in modular social networks \cite{BanerjeeFudenbergWordofMouth} or propagating response-driven surveys across bottlenecks between social groups \cite{SalganikHeckathornRDS}. Cascades in social networks like these may require networks with triangles or other subgraphs added \cite{newman_subgraphs, miller_percolation, newman_clustering}; inverting the resulting multidimensional generating function equations for dynamics on these networks would require similar multitype techniques.

We expect that this work will stimulate calculations of critical points in interconnectivity among networks subjected to other dynamics, such as linearized power flow equations in electrical grids \cite{Pepyne2007,HinesChaos2010} and other domain-specific models. As critical infrastructures such as power grids, transportation, communication, banks and markets become increasingly interdependent, resolving the risks of large cascades and the incentives that shape them becomes ever more important.

\section{Materials and Methods}
\subsection{Power grid topologies}\ To understand coupling between multiple grids we requested data from the US Federal Energy Regulation Commission~\cite{FERC}. Using output of power simulations on many connected ``areas'' (distinct grids owned by different utilities) in the southeastern USA, we chose grid $d$ by selecting the grid with the highest average internal degree and the grid, $c$, to which it had the most interconnections (8). Grids $c,d$ have 439 and 504 buses (nodes) and average internal degrees 2.40 and 2.91. For our purposes here, the only important details are the narrow degree distributions, the low clustering coefficients and the number of interconnections between the grids. Other details about the grids are in the SI.

\subsection{Toppling branch distributions}\ 
To reduce the number of types in the branching process, we count the number of toppling events in each network rather than the number of grains of sand shed from one network to another. Here we derive the toppling branch distributions $u_a, u_b$ from the shedding branch distributions $q_{od}$. (For instance, $u_a(t_a,t_b)$ is the chance that a toppled node in $a$ causes $t_a, t_b$ many nodes in $a,b$ to topple at the next time step.) Note that a node topples if and only if it sheds at least one grain of sand. Thus a grain traveling from a network $o$ to network $d$ topples its destination node with probability $1-q_{od}(0,0)$. Denoting $(0,0)$ by $\mathbf{0}$ and the Binomial distribution by $\beta_k^n(p) \equiv \binom{n}{k}p^{k} (1-p)^{n-k}$, we have 
\begin{align*}
u_a(t_a, t_b) = \sum_{k_a = t_a}^\infty \sum_{k_b = t_b}^\infty p_a(k_a, k_b) & \beta_{k_a}^{t_a}(1-q_{aa}(\mathbf{0}))  \beta_{t_b}^{k_b}(1-q_{ab}(\mathbf{0})),\notag 
\end{align*} 
since the node must have at least $t_a$ many $a$-neighbors and at least $t_b$ many $b$-neighbors, only $t_a, t_b$ of which topple (which are binomially distributed). The expression for $u_b$ is analogous.  As an example, the probability generating function of $u_a$ for Bernoulli-coupled random regular graphs \RBR{z_a}{p}{z_b} is
\small
\begin{align*} 
\mathcal{U}_a(\tau_a, \tau_b) = \frac{(p-p \tau_a+(z_a+1)(\tau_a+z_a-1))^{z_a}(1+p(\tau_b-1)+z_b)}{(z_a+1)^{z_a} z_a^{z_a}(z_b+1)}.
\end{align*}
\normalsize

\subsection{Three methods for numerically solving the multidimensional self-consistency equations of a multitype branching process}\ 
The most na\"ive method to solve the self-consistency equations of a multitype branching process (such as Eqs. \eqref{selfconsistencyequations}) is to use a computer algebra system like \emph{Mathematica} or \emph{Maple} to iterate \eqref{selfconsistencyequations} symbolically starting from $\mathcal{S}_a = \mathcal{S}_b = 1$, expand the result, and collect coefficients. To obtain the coefficient $s_a(t_a, t_b)$ exactly, it suffices to iterate \eqref{selfconsistencyequations} at least $t_a + t_b + 1$ times. What is more, just a handful of iterations partially computes the tails (coefficients of terms with high powers in $\mathcal{S}_a, \mathcal{S}_b$), but the amount of missing probability mass in the tails is undetermined. This method was used in Fig.~\ref{matchtheorysimulation_f1}.

A second method is to symbolically iterate Eqs. \eqref{selfconsistencyequations} at least $t_a + t_b + 1$ times and to use Cauchy's integral formula
\begin{align}\label{Cauchyintegral}
s_a(t_a, t_b) = \frac{1}{(2 \pi i)^2} \int \hspace{-2mm} \int_D \, \! \frac{\mathcal{S}_a(\tau_a, \tau_b)}{\tau_a^{t_a+1}\tau_b^{t_b+1}} \, d\tau_a d\tau_b,
\end{align}
where $D \subset \mathbb{C}^2$ is a Cartesian product of circular contours centered at the origin, each of radius $r$ smaller than the modulus of the pole of $\mathcal{S}_a$ closest to the origin \cite{Good_Lagrange}. 
 Then calculate one coefficient at a time using 
\begin{align*}
s_a(t_a, t_b) = \frac{1}{4 \pi^2} \int_0^{2 \pi} \hspace{-1.5mm} \int_0^{2 \pi}  \frac{\mathcal{S}_a(r e^{i \theta}, r e^{i \psi})}{r^{t_a+t_b} e^{i(t_a \theta + t_b \psi)}} \, d\theta d\psi.
\end{align*}

A third method is Lagrange inversion, generalized to multiple dimensions by I. J. Good in 1960 \cite{Good_Lagrange}, a result that has seen little use in the networks literature. We state the theorem in the language of the two-type branching process considered here, with the notation $\vec \tau \equiv (\tau_a, \tau_b)$, $\vec{\mathcal{S}} \equiv (\mathcal{S}_a, \mathcal{S}_b)$ and $\vec{\mathcal{U}} \equiv (\mathcal{U}_a, \mathcal{U}_b)$.
For the slightly more general result that holds for arbitrary, finite, initial population, see Ref. \cite{Good_Lagrange}.
\begin{theo}\label{LagrangeTheorem}[Good 1960] If $\vec{\mathcal{U}}(\vec \tau)$ is analytic in a neighborhood of the origin and $\vec{\mathcal{U}}(\vec \tau) \not = 0$ (i.e., every type has a positive probability of being barren), then for $x \in \{0,1\}$,
\begin{align}\label{eq:theorem10}
\mathcal{S}_a(\vec \tau)^x \mathcal{S}_b(\vec \tau)^{1-x} =&  \sum_{m_a, m_b = 0}^\infty   \frac{\tau_a^{m_a} \tau_b^{m_b}}{m_a! m_b!} \bigg [ \frac{\partial^{m_a+m_b}}{\partial \kappa_a^{m_a} \partial \kappa_b^{m_b}} \bigg \{ \kappa_a^x \kappa_b^{1-x} \times  \\& \times \mathcal{U}_a(\vec \kappa)^{m_a}  \mathcal{U}_b(\vec \kappa)^{m_b}             \bigg | \bigg | \delta_\mu^\nu - \frac {\kappa_\mu}{\mathcal{U}_\mu} \frac {\partial \mathcal{U}_\mu}{\partial \kappa_{\mu}} \bigg | \bigg |            \bigg \} \bigg ]_{\vec \kappa = 0}, \notag
\end{align}
where $\mu, \nu$ run over the types $\{a,b\}$, $\delta_\mu^\nu$ is the Kronecker delta, and $|| \cdot ||$ is the determinant.\label{theorem10}\end{theo}
Taking $x = 0$ gives $\mathcal{S}_b$, while taking $x=1$ gives $\mathcal{S}_a$. In practice, hundreds of terms of $\mathcal{S}_a,\mathcal{S}_b$ can be computed exactly by truncating the sum in \eqref{eq:theorem10} and using computer algebra systems.

\subsection{Solving for the coefficients asymptotically (and why standard techniques fail)}\ To determine the asymptotic behaviors of $s_a(t_a, t_b), s_b(t_a, t_b)$ as $t_a, t_b \rightarrow \infty$, the trick is to solve for the inverses of $\mathcal{S}_a$ and $\mathcal{S}_b$ and to expand those inverses $\mathcal{S}_a^{-1}$ and $\mathcal{S}_b^{-1}$ at the singularities of $\mathcal{S}_a$ and $\mathcal{S}_b$. This technique works for isolated networks \cite{Goh_sandpile_PRL,Goh_sandpile_PhysicaA, Lee:2005fo, watts_newman_strogatz}, but $\mathcal{S}_a$ and $\mathcal{S}_b$ have no finite singularities in $\mathbb{C}^2$ for Bernoulli-coupled random regular graphs nor for power grids $c,d$.

We demonstrate this failure of standard asymptotic techniques on the networks \RBR{z_a}{p}{z_b}. Let $\vec \omega := \vec{\mathcal{S}}(\vec \tau)$. Assuming an inverse $\vec{\mathcal{S}}^{-1}$ of $\vec{\mathcal{S}}$ exists, using \eqref{selfconsistencyequations} we have
\begin{align}\label{Sinverse}
\vec \tau = \vec{\mathcal{S}}^{-1}(\vec \omega) = \frac{\vec \omega}{\vec{\mathcal{U}}(\vec \omega)} = \bigg (\frac{\omega_a}{\mathcal{U}_a(\vec \omega)}, \frac{\omega_b}{\mathcal{U}_b(\vec \omega)} \bigg ).
\end{align}
The generating functions $\mathcal{S}_a, \mathcal{S}_b$ have singularities at $\vec \tau^*, \vec \mu^*$, respectively, if $\vec D \mathcal{S}_a^{-1}(\vec \tau^*) = \vec D \mathcal{S}_b^{-1}(\vec \mu^*) = (0,0) $, where the operator $\vec D = (\partial / \partial \omega_a, \partial / \partial \omega_b)$ is a vector of partial derivatives. Differentiating Eq. \eqref{Sinverse} and equating the numerators to 0 gives
\begin{subequations} \label{derivatives}
\begin{align}
\mathcal{U}_a(\vec \tau^*) - \tau_a^* \frac{\partial \mathcal{U}_a}{\partial \tau_a}(\vec \tau^*) = 0, \quad -\tau_a^*  \frac{\partial \mathcal{U}_a}{\partial \tau_b}(\vec \tau^*) = 0,\\
\mathcal{U}_b(\vec \mu^*) - \mu_b^* \frac{\partial \mathcal{U}_b}{\partial \tau_b}(\vec \mu^*) = 0, \quad -\mu_b^*  \frac{\partial \mathcal{U}_b}{\partial \tau_b}(\vec \mu^*) = 0.
\end{align}
\end{subequations}
The only solution to Eqs. \eqref{derivatives} is
\begin{subequations}\label{badsolution}
\begin{align}
\vec \tau^*_1 &= \bigg (\frac{z_a^2 + p - 1}{p-z_a-1}, \frac{p-z_b-1}{p} \bigg ) \xrightarrow[p \rightarrow 0]{} (\infty, 1-z_a),\\
\vec \mu^*_1 &= \bigg (\frac{p-z_a-1}{p}, \frac{z_b^2 + p - 1}{p-z_b-1} \bigg )  \xrightarrow[p \rightarrow 0]{} (1-z_b, \infty).
\end{align}
\end{subequations}
This solution \eqref{badsolution} does not recover the singularities $\vec \tau^* = (1,c_1), \vec \mu^* = (c_2,1)$ (where $c_1, c_2$ are arbitrary) that we should obtain when the networks are isolated ($p=0$) \cite{Goh_sandpile_PRL}. Moreover, although Eqs.~\eqref{derivatives} vanish at the solutions \eqref{badsolution}, the derivatives of \eqref{Sinverse} do not vanish at these solutions \eqref{badsolution}, as $\vec D \mathcal{S}_a^{-1}(\vec \tau^*_1) = (\infty, 0)$, $\vec D \mathcal{S}_b^{-1}(\vec \mu^*_1) = (0, \infty)$. Thus we must discard solutions \eqref{badsolution}.

Solving only the left-hand equations in Eqs. \eqref{derivatives} yields singularities that do recover the correct singularities when $p=0$:
\begin{subequations}\label{bettersolution}
\begin{align}
\vec \tau^*_2 &=\bigg (\frac{z_a^2 + p - 1}{(1-z_a)(p-1-z_a)}, \overline{\tau_b} \bigg ) \xrightarrow[p \rightarrow 0]{} (1, \overline{\tau_b}),\\
\vec \mu^*_2 &= \bigg (\overline{\mu_a}, \frac{z_b^2 + p - 1}{(1-z_b)(p-1-z_b)} \bigg ) \xrightarrow[p \rightarrow 0]{} (\overline{\mu_a}, 1),
\end{align}
\end{subequations}
where $\overline{\tau_b}, \overline{\mu_a}$ are arbitrary constants satisfying
\begin{align}
\overline{\tau_b} \neq \frac{p - z_b - 1}{p}, \quad \overline{\mu_a} \neq \frac{p - z_a - 1}{p},
\end{align}
so that the derivatives of $\vec{\mathcal{S}}^{-1}(\vec \omega)$ \eqref{derivatives} evaluated at \eqref{bettersolution} are finite.
However, the derivatives on the right-hand sides of Eqs. \eqref{derivatives} evaluated at the solutions \eqref{bettersolution} are
\begin{align*}
\frac {\partial \mathcal{S}_a^{-1}}{\partial \omega_b}(\tau_2^*) = \bigg (0, \frac{p \left(z_b+1\right) \left(z_a^2+p-1\right) \left(\frac{z_a^2-1}{z_a^2+p-1}\right)^{z_a}}{\left(1-z_a\right) \left(1-p+z_a\right) \left(p \left(\omega _b-1\right)+z_b+1\right)^2} \bigg)
\end{align*}
and similarly for $\partial \mathcal{S}_b^{-1} / \partial \omega_a(\mu_2^*)$ (interchange $a$ and $b$). This derivative vanishes if and only if $p = 0$ or $p = 1-z_a^2$ or $z_a=1$ or $\omega_b = \infty$. Hence we cannot find finite singularities of $\mathcal{S}_a, \mathcal{S}_b$; these generating functions are entire functions with singularities only at $\tau_b = \mu_a = \infty$, $\tau_a \neq \frac{z_a^2 + p - 1}{p-z_a-1}, \mu_b \neq \frac{z_b^2 + p - 1}{p-z_b-1}$. Hence we have no singularities at which to do an asymptotic expansion for the coefficients $s_a, s_b$, as one can for isolated networks \cite{Goh_sandpile_PRL, Goh_sandpile_PhysicaA, Lee:2005fo, watts_newman_strogatz}.

Other techniques exist for asymptotically approximating the coefficients of generating functions, depending on the type of the singularity (\cite{Wilf} \S 5). Hayman's method (\cite{Wilf} \S 5.4) works for generating functions with no singularities in the finite plane (i.e., entire functions), like the $\mathcal{S}_a, \mathcal{S}_b$ considered here. However, the theorem requires a closed form expression for the generating function, which we cannot obtain from the self-consistency equations for the synthetic and real interacting networks of interest. Developing techniques for asymptotically approximating the coefficients of multidimensional generating functions with singularities at infinity, like those studied here, poses an important challenge for future studies of dynamical processes on multitype networks.

\noindent
\section{Appendix: Effective degree distributions in multitype networks}
Using the configuration model to generate multitype networks---including bipartite and multipartite graphs, graphs with arbitrary distributions of subgraphs, and the modular graphs considered here---requires matching edge stubs within and among types. For example, for the interacting networks considered here, the number of edge stubs pointing from network $a$ to network $b$ must equal the number from $b$ to $a$. A standard practice in the literature \cite{PourboPRE09,newman_subgraphs} that is more restrictive than needed is to require that the averages of the inter-degrees over the degree distributions agree (e.g., $\langle k_{ab} \rangle = \langle k_{ba} \rangle$ for two networks $a,b$ of equal size). In fact, most any degree distributions can be used, as long as conditioning on matching edge stubs among the types of nodes leaves some probability. Requiring that the degree distributions satisfy, for example, $\langle k_{ab} \rangle = \langle k_{ba} \rangle$, merely tips the scales in favor of valid degree sequences.

Here we derive the effective degree distribution in multitype networks generated with the configuration model. The idea is simple: since nodes draw degrees independently, the probability distribution of the total number of edge stubs from, say, network $a$ to $b$ is given by a convolution of the degree distribution. We state it for the two interacting networks considered here, but it can be easily generalized to, say, role distributions~\cite{newman_subgraphs}, which require matching edge stubs with ratios different from one.

Suppose networks $a,b$ have $N_a, N_b$ many nodes, respectively. Let $\vec K_{ab}, \vec K_{ba}$ be the random variables for the sequences of ``inter-degrees''  $k_{ab}, k_{ba}$ of the nodes in networks $a, b$, respectively, drawn from the input degree distributions $p_a(k_{aa}, k_{ab}), p_b(k_{ba},k_{bb})$. Suppose, for simplicity, that the internal and external degrees are independent ($p_a(k_{aa}, k_{ab}) = p_{aa}(k_{aa}) p_{ab}(k_{ab})$, $p_b(k_{ba}, k_{bb}) = p_{ba}(k_{ba}) p_{bb}(k_{bb})$), although this is not essential. We denote $\Sigma \vec k \equiv \sum_{i=1}^n k_i$ for $\vec k \in \mathbb{Z}^n$.
\begin{lem}\label{inter_degree_lemma} With the assumptions in the previous paragraph, the \emph{effective inter-degree distribution} for network $a$ is \emph{not} the input one, $\Pr(\vec K_{ab} = \vec k) = \prod_{i=1}^{N_a} p_{ab}(k_i)$, but rather the conditional probability distribution
\begin{align*}
\Pr(\vec K_{ab} = \vec k \mid \Sigma \vec{K}_{ba} = \Sigma  \vec K_{ab}) = \Pr(\vec K_{ab} = \vec k) \, \frac {p_{ba}^{*}\big ( \Sigma \vec k \big)} {\sum_{\ell} p_{ab}^{*} (\ell)p_{ba}^{*}(\ell)},
\end{align*}
where $p_{ab}^{*}(\cdot), p_{ba}^{*}(\cdot)$ are $p_{ab}, p_{ba}$ convolved $N_a, N_b$ times, respectively.\end{lem}
\noindent \textsc{Proof. } Using the independence of $\vec K_{ab}$ and $\vec K_{ba}$, we have
\begin{align}
\Pr( \vec K_{ab} = \vec k & \mid \Sigma \vec K_{ba} = \Sigma \vec K_{ab})= \frac{\Pr(\vec K_{ab} = \vec k)  \Pr(\Sigma \vec K_{ba} = \Sigma \vec k)}{\Pr(\Sigma \vec K_{ba} = \Sigma \vec K_{ab})}. \notag 
\end{align}
In the numerator we recognize $\Pr(\Sigma \vec K_{ba} = \Sigma \vec k)$ to be $p_{ba}$ convolved $N_b$ times, evaluated at $\Sigma \vec k$. In the denominator, use independence, recognize convolutions and sum on $\ell \geq 0$.\hfill $\Box$

Lemma \ref{inter_degree_lemma} shows that the \emph{effective} inter-degree distribution of $a$ is the input degree distribution $p_{ab}$ reduced by an amount given by a fraction of convolutions. This reduction in probability governs how many invalid degree sequences must be generated before generating a valid one. For systems sizes on the order of $10^4$ nodes, as considered here, generating degree sequences until producing a valid one is quite feasible, as it takes merely seconds. However, for millions of nodes or more, it is better to generate degree sequences $\vec K_{ab}, \vec K_{ba}$ once, and then repeatedly choose a node uniformly at random to redraw its degree from its degree distribution until the degree sequences are valid. However, this method does not escape the effect in Lemma \ref{inter_degree_lemma}, which is often subtle but can be substantial if the supports of the convolutions of the degree distributions have little overlap. The inter-degree distributions used here, Bernoulli and Correlated-Bernoulli with identical expected total inter-degree, have ``much overlap'', so the effective inter-degree distribution is approximately the input one, and the correction factor in Lemma \ref{inter_degree_lemma} can be neglected.

\begin{acknowledgments}
C.D.B. was partially supported by NSF VIGRE DMS0636297, a Graduate Assistance in Areas of National Need (GAANN) Fellowship, the Statistical and Applied Mathematical Sciences Institute (SAMSI), and the Department of Defense (DoD) through the National Defense Science \& Engineering Graduate Fellowship (NDSEG) Program. We also gratefully acknowledge support from the following agencies: the Defense Threat Reduction Agency, Basic Research Award DTRA1-10-1-0088, the Army Research Laboratory, Cooperative Agreement Number W911NF-09-2-0053, and the National Academies Keck Futures Initiative, Grant No. CS05. Finally we thank the members of the SAMSI Dynamics on Networks working group for useful discussions. 
\end{acknowledgments}



\renewcommand{\thesection}{S\arabic{section}}
\renewcommand{\thesubsection}{\thesection.\arabic{subsection}}
\renewcommand{\theequation}{S\arabic{equation}}
\renewcommand{\thefigure}{S\arabic{figure}}
\renewcommand{\thetable}{S\arabic{table}}


\setcounter{section}{0}
\setcounter{figure}{0}

\begin{center}
{\bf Supporting information}
\end{center}

\section{Robustness of optimal interconnectivity} 
\subsection{How $p^*$ depends on system size, connectivity, type of degree distribution}\ Our central result is the minimum $p^*>0$ in the chance of a large cascade in a network with a fraction $p$ of its nodes connected to nodes in another network. Important questions are how $p^*$ depends on system size, behaves in the thermodynamic limit (infinite system size), and depends on the internal degrees and on the degree distribution. Here we show that the qualitative form of $\Pr(T_a> \cutoff)$ in Fig. 4 appears to be generic.

\begin{figure}[hbt]
\begin{center}
\centerline{\includegraphics[width=8.7cm]{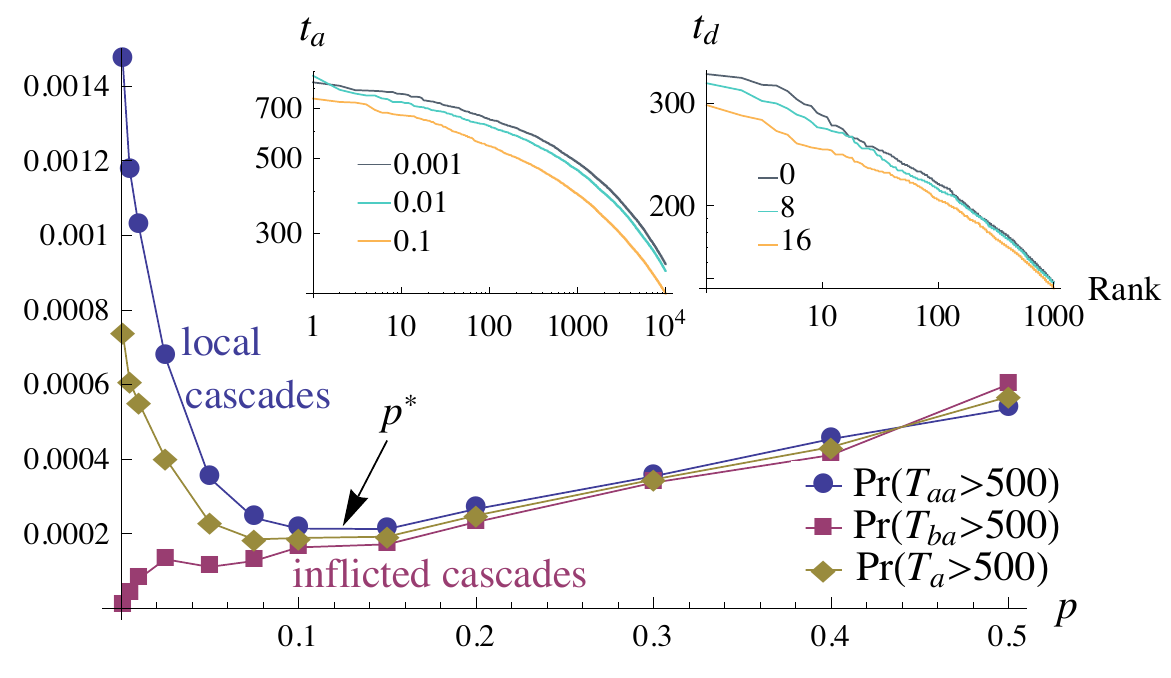}}
\caption{For half the system size ($N_a=1000$), half the cutoff ($\cutoff=500$) and double the dissipation ($f=0.02$) 
of the system in Fig.~4, the chance of a large cascade in $a$ is qualitatively similar, and $p^* \approx 0.12 \pm 0.02$ is slightly larger (in simulations with $2\times10^6$ grains dropped after $10^5$ transients). This gold curve and its critical point is stable to cutoffs $\cutoff$ different from 500 ($200 \leq \cutoff \leq 800$). Inset figures: rank-size plots on log-log scales of the largest cascades in network $a$ (left) for $p=10^{-3}, 10^{-2}, 10^{-1}$ and in power grid $d$ connected to $c$ by 0, 8 or 16 edges.}
\label{locallystabilizing-halfN}
\end{center}
\end{figure}

Doubling the system size and keeping the cutoff $\cutoff$ and the dissipation $f$ fixed (at $\cutoff = $ half the number nodes in $a$ and at $f=0.01$) does not significantly change Fig.~4, because the dissipation $f$ limits the chance of large cascades. But doubling the system size, doubling the cutoff $\cutoff$ and halving the dissipation $f$ (so that the largest cascades topple nearly the whole system) slightly decreases $p^*$. Figure~\ref{locallystabilizing-halfN} shows similar results as Fig.~4 but for a system half the size ($1000$ nodes/network, $f=0.02$, cutoff $\cutoff=500$). In the thermodynamic limit of infinite system size $N$, we expect $p^*$ to stay bounded away from 0, because what appears to be the determinant is the ratio of edges between the networks ($pN$) and within them ($z_a N$).

\begin{figure}[hbt]
\begin{center}
\centerline{\includegraphics[width=8.7cm]{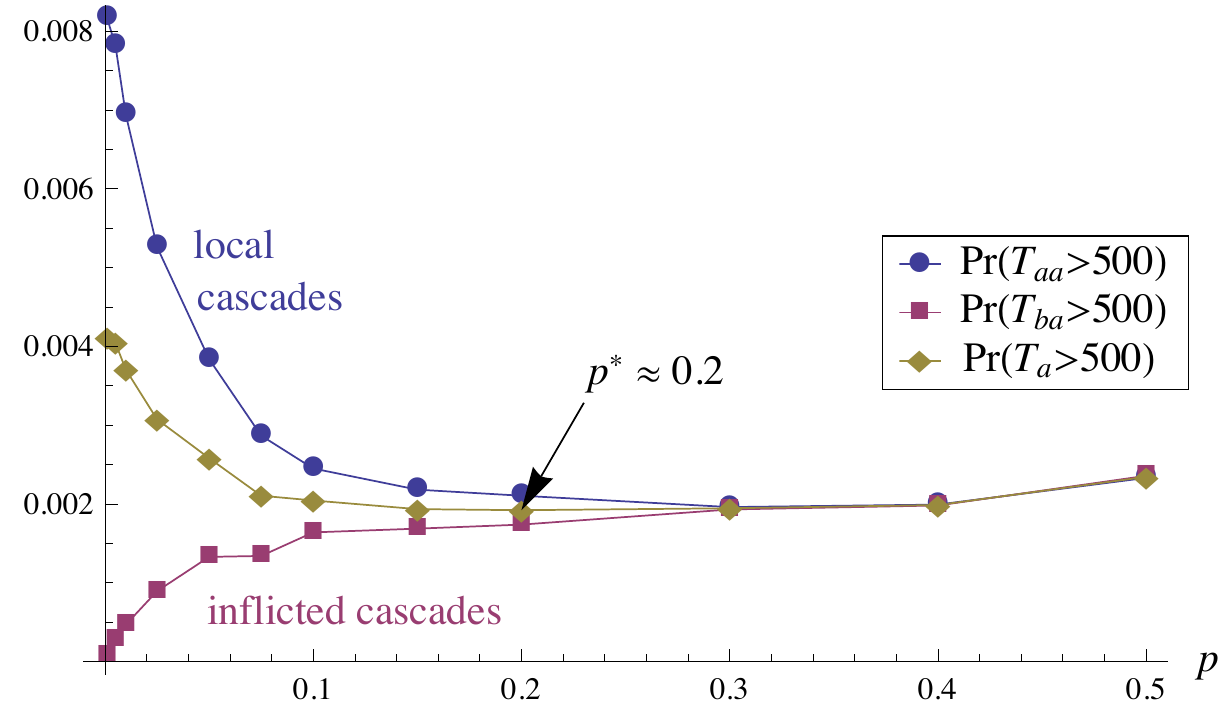}}
\caption{Increasing the internal degrees of both networks $a,b$ from 3 to 4 (i.e., \RBR{4}{p}{4}) slightly increases the minimum $p^*\approx 0.2$ of the chance of a large cascade in network $a$. (Here: $10^3$ nodes/network; $f=0.02$; $10^6$ grains dropped after $10^5$ transients.)}
\label{fig:R4}
\end{center}
\end{figure}

Increasing the internal degrees of nodes (say, to coupled random 4-regular graphs \RBR{4}{p}{4}) increases their capacity and hence ability to withstand inflicted cascades, so the minimum in $\Pr(T_a>\cutoff)$ is therefore larger, $p^*\approx 0.2$ (Fig.~\ref{fig:R4}), compared to $p^* \approx 0.12$ for random 3-regular graphs with equally many nodes (Fig.~\ref{locallystabilizing-halfN}). The networks also have a wider range of optimal $p$, as expected, given that $pN_a$ external edges is less significant than the $z_a N_a$ internal edges when $z_a = 4$ (Fig.~\ref{fig:R4}) compared to $z_a = 3$ (Figs.~4 and~\ref{locallystabilizing-halfN}). Nonetheless, the chance of a large cascade eventually increases with $p$. We also note that power grids more closely resemble random 3-regular than 4-regular graphs (see Table \ref{tab:statistics} below).

Next we determined the effect of introducing some degree heterogeneity by adjusting interconnectivity $p$ between two \ER random graphs. Similar results, were obtained as for random regular graphs indicating that some degree heterogeneity does not affect these results. (We did not test heavy-tailed degree distributions, since these rarely occur in the physical infrastructure networks of interest, namely power grids.)

Finally, we tested whether adding edges at random to a random regular graph produces an optimal connectivity that minimizes the chance of a large cascade.  Specifically, each node in a 3-random regular graph gains an extra, fourth edge stub independently with probability $p$ (i.e., degree distribution $P(k)$ is $P(3)=1-p,P(4)=p$). We find in simulations that $Pr(T>500)$ for a system of $10^3$ nodes is largely independent of the mean degree $z$. In particular, the chance of a large cascade does not drop, as it does when initially adding edges between two networks (Figs.~4,~\ref{locallystabilizing-halfN},~\ref{fig:R4}). This agrees with theoretical results for isolated networks~\cite{Goh_sandpile_PRL}: the avalanche size distribution $s(t) \sim t^{-3/2}$ for large $t$. {\it We conclude that adding links between two networks affects vulnerability to large cascades in a different way than adding links within isolated networks.}

\subsection{Unstable $p^*$ for small cascades, stable $p^*$ for large cascades}
\begin{figure}[htbp]
\begin{center}
\includegraphics[width=8.7cm]{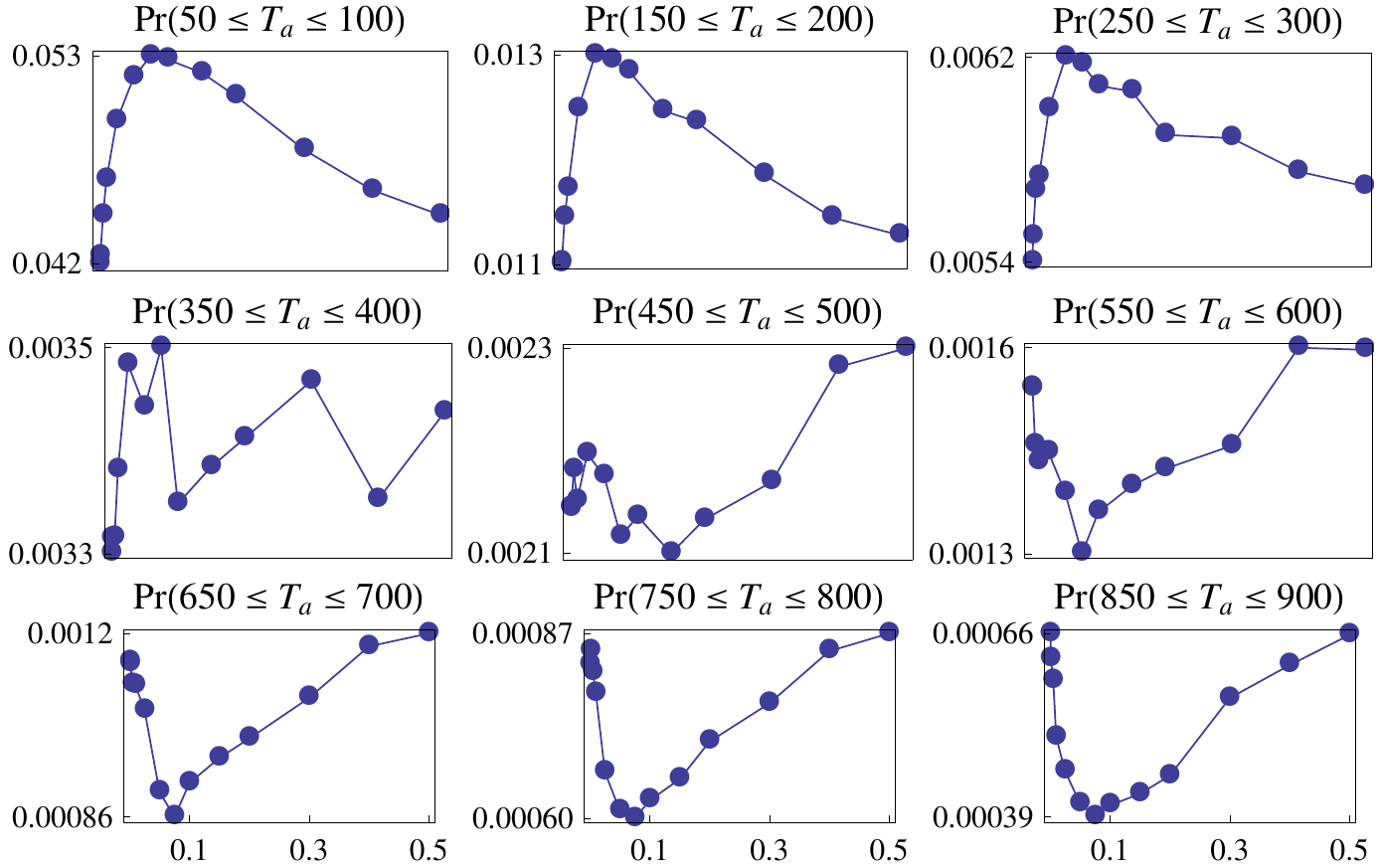}
\caption{Plots of $\Pr(\ell \leq T_a \leq \ell + 50)$, for $\ell = 50, 150, ..., 850$, as a function of interconnectivity $p \in [0, 0.5]$, in simulations on \RBR{3}{p}{3} with $2\times 10^6$ grains, $2\times10^3$ nodes/network, $f=0.01$. Intermediate cascades ($50 \leq T_a \leq 300$) have an unstable critical point $p^* \approx 0.05$, which changes concavity for cascades of size approximately 350. Large cascades $T_a \geq 500$ have an stable critical point $p^* \approx 0.075$.}
\label{moviereel}
\end{center}
\end{figure}

\ 
Networks seeking to mitigate cascades of small, intermediate or large sizes would seek different interconnectivity, as shown in the plots of $\Pr(\ell \leq T_a \leq \ell+50)$ as a function of $p$ in Fig.~6. In Fig.~\ref{moviereel} we show more plots to show the change in concavity at intermediate cascade size and the robustness of the location of the stable minimum $p^* \approx 0.075$ for \RBR{3}{p}{3}. 

\subsection{Increasing capacity fuels larger system-wide cascades}\label{subset:globalcascades}\ 
Introducing new interconnections between networks slightly enlarges the largest global cascades. We ask to what extent is this due to the direction of links (pointing internally or externally) and due to the additional capacity for holding sand endowed by the new links (since capacities of nodes are their degrees). To isolate these two effects, we run simulations on two random regular graphs interconnected by ``Correlated-Bernoulli'' coupling: each node changes an internal edge stub to an external edge stub with probability $p$.

\begin{figure}[hbt]
\begin{center}
\centerline{\includegraphics[width=8.7cm]{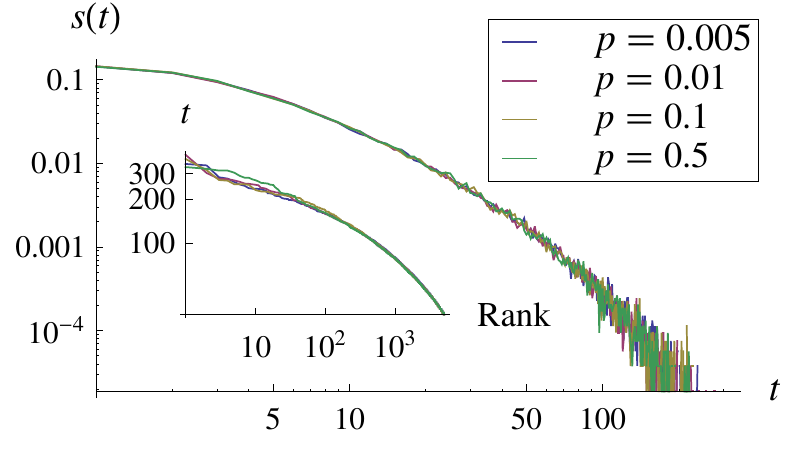}}
\label{correlated_totalavalanchesize}
\caption{When nodes change exactly one of their internal edge stubs to be an external edge stub with  probability $p$ (Regular($z_a$)-Correlated-Bernoulli($p$)-Regular($z_b$) rather than receiving an external edge stub with probability $p$ (Regular($z_a$)-Bernoulli($p$)-Regular($z_b$)), global cascades are not significantly amplified. This indicates that it is the small increase in capacity (due to the additional edges and the fact that capacities of nodes are their degrees) rather than the direction of edges (internal versus external) that causes the increase in global cascades with increased Bernoulli-distributed coupling between networks. The main plot is the total avalanche size distribution $s(t)$ in a simulation with $10^6$ grains of sand dropped (after $10^5$ grains dropped without collecting statistics, begun from initial amounts of sand chosen uniformly at random from zero to one less a node's degree) on two networks with $10^3$ nodes each, with dissipation of sand $f=0.05$ and $z_a=z_b=3$. \emph{Inset}: rank-size plot of largest $5\times10^3$ cascades, which are nearly indistinguishable.}
\end{center}
\end{figure}

The resulting total avalanche size distribution, Fig.~\ref{correlated_totalavalanchesize}, does not significantly change for different values of interconnectivity $p$. Thus, what causes the slight enlargement of the largest global avalanches is the slight increase in total capacity of the network, not the direction of links. Moreover, the enlargement of the largest global cascades is consistent with the increase in capacity. For the simulations on random 3-regular graphs in Fig.~\ref{correlated_totalavalanchesize}, the $n$th largest avalanche contains on average $1.4 \pm 2.6$  (mean $\pm$ standard deviation) more topplings with $p=.1$ compared to $p=.005$, which is an insignificant difference. By contrast, in an analogous simulation on \RBR{3}{0.1}{3}, in which we introduce extra links with probability $p$, the $n$th largest avalanche contains $204.8 \pm 92.5$ more topplings than \RBR{3}{0.005}{3}, a significant difference for networks with $10^3$ nodes. Furthermore, this difference is on the order of the additional capacity of $2 \times 10^3 \times (0.1 - 0.005) = 190$ among $2\times10^3$ nodes with $B$($.1$) compared to $B$($.005$) coupling.

We find similar results when introducing eight additional interconnections between power grids $c$ and $d$. The $n$th largest global cascades in simulations are larger by an amount on the order of the additional capacity of the two networks. Thus the additional capacity due to the new interconnections, not the direction of links, largely explains the amplification of system-wide cascades.

\subsection{Bounding the critical load disparity $r^*$ for power grids $c,d$}\ 
In the main paper, we showed that for 16 interconnections between power grids $c$ and $d$, the critical load disparity is $r^* \approx 15$. That is, when sand is dropped 15 times more frequently on $c$-nodes than on $d$-nodes, the largest inflicted cascades from $c$ to $d$ approximately equal in size the largest local cascades begun in $d$. (Recall that by ``inflicted cascade from $c$ to $d$'' we mean the number of topplings in $d$ in an avalanche begun in $c$, and by ``local cascade in $d$'' we mean the number of topplings in $d$ in an avalanche begun in $d$.) If $d$ builds more interconnections (or increases the load disparity $r$), the largest inflicted cascades become larger than the largest local cascades. On the other hand, delete interconnections (or decrease the load disparity $r$), and $d$ could mitigate its largest local cascades more than the enlargement of the largest inflicted cascades. Hence $r^*=15$ and 16 interconnections represent a ``modularity equilibrium'' in local and inflicted cascades.

\begin{figure}[hbt]
\begin{center}
\centerline{\includegraphics[width=8.7cm]{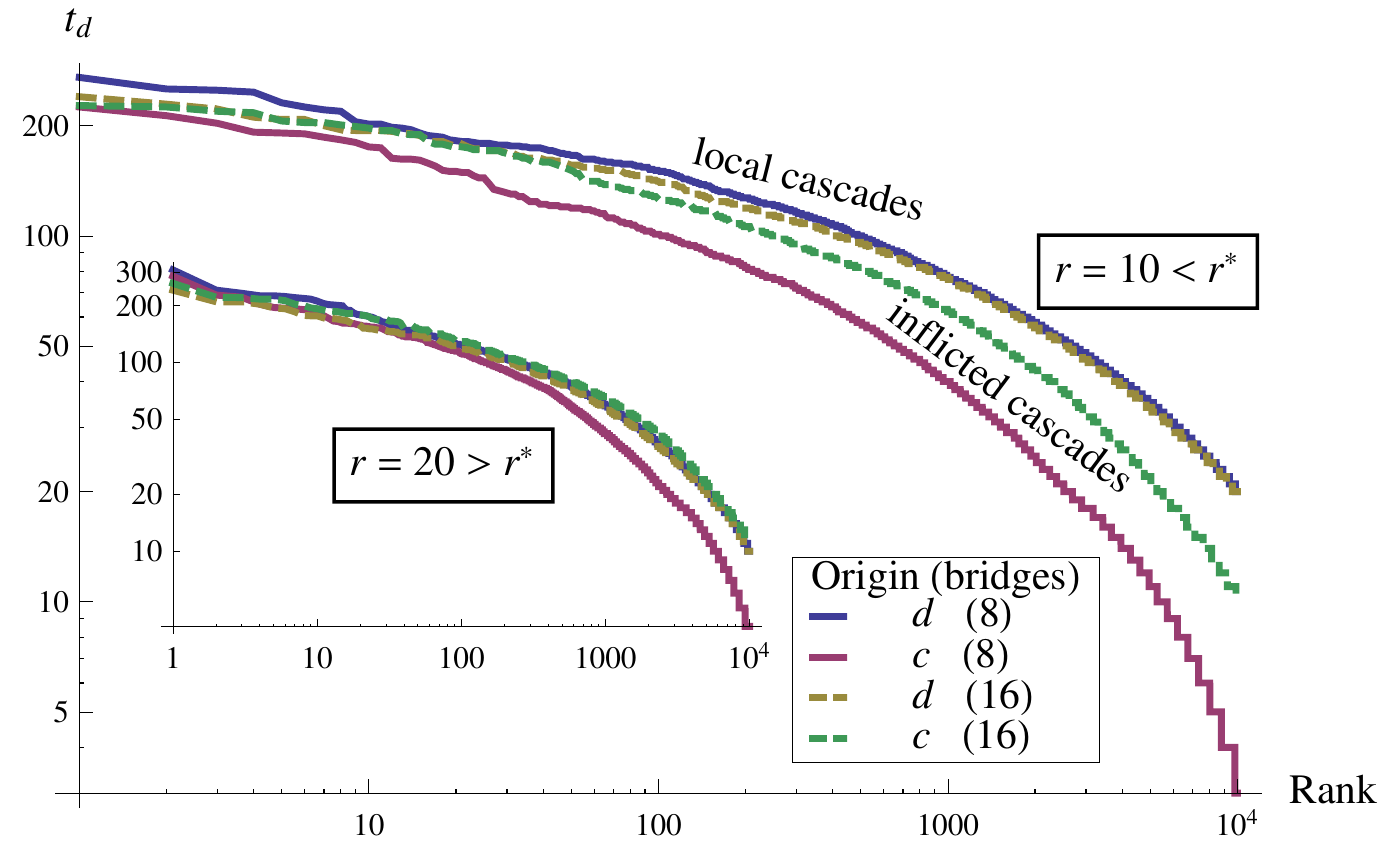}}
\label{rankplot_cd_r10andr20}
\caption{Same plot as in Fig.~7, but for load disparity $r = 10$ (main plot) and $r=20$ (inset). For 16 edges between grids $c$ and $d$ (dashed curves), the largest inflicted cascades from grid $c$ to $d$ (green dashed curve) are slightly smaller than the largest local cascades from $d$ to itself (gold dashed curve) for $r=10$, indicating that $r^* \gtrapprox 10$. For $r=20$ (inset), the largest inflicted cascades are slightly larger than the largest local cascades, indicating that $r^* \lessapprox 20$.}
\end{center}
\end{figure}

Here we use simulations to approximately bound the critical load disparity $10 \lessapprox r^* \lessapprox 20$ for 16 interconnections between $c$ and $d$. As shown in Fig.~\ref{rankplot_cd_r10andr20}, for $r=10, 20$ the largest inflicted cascades (green dashed) are smaller and larger, respectively, than the largest local cascades (gold dashed), indicating that $10 \lessapprox r^* \lessapprox 20$ for 16 interconnections. For 8 interconnections (solid blue and red curves), the critical load disparity $r^*$ is evidently larger than 20, as the largest inflicted cascades are smaller than the largest local cascades for $r=20$ (red curve below blue curve).

\section{Power Grid Topologies }
\subsection{Power grids $c$ and $d$}\ 
The readily available datasets on power grids---IEEE test cases \cite{WashingtonData}, Western States \cite{Watts1998}, and NYISO \cite{Wang2010}---did not appear to contain multiple electrical grids connected to one another.  As a result, we requested data on power grid connectivity from the US Federal Energy Regulation Commission (FERC) via the Critical Energy Infrastructure Information program \cite{FERC_CEII}. We focused on the southeastern region of the United States, for which we had the output files of power simulations of various grids. The southeastern region consists of \emph{areas}, distinct grids owned by different utility companies. The largest areas contain thousands of buses, an electrical grid term for the connection points that link generators, transmission lines and transformers. We ignore wind turbines, for they do not belong to specific areas. Among the areas with at least 100 nodes, we chose grids $c$ and $d$ by first selecting the grid with the highest average internal degree (2.91), and then selecting the area, $c$, to which it had the most interconnections (8 of them). Grid $d$ consists of one giant component with 504 nodes, while grid $c$ consists of one giant component of size 439, and we ignore 14 additional nodes.

Two of the statistics of power grids $c$ and $d$ are important to our study: the narrow degree distribution and small clustering coefficient.  Some details are in Table \ref{tab:statistics}. The average clustering coefficient $\langle C \rangle$, the fraction of possible triangles that exist, begins to measure how locally tree-like power grids are. We find that $\langle C \rangle \approx 0.05$ is low, yet an order of magnitude larger than the clustering coefficient of a random 3-regular graph with as many nodes as grid $d$. The average shortest path length among pairs of nodes, $\langle \ell \rangle$, which has been identified as a likely source of difficulty for predicting cascades on networks \cite{MelnikUnreasonable2011}, is rather large in the power grids ($\langle \ell \rangle \approx 10$), due to their two-dimensional, spatial, nearly planar structure. By contrast, random 3-regular graphs have significantly smaller diameter $\langle \ell \rangle \approx 7.09\pm 0.03$.

\begin{table}[hbt]
 \label{tab:statistics}
\centering
\begin{tabular}{ c |  c c c c }
      & $c$ & $d$ & $c$ \& $d$ & 3-Regular \\ \hline
\# nodes & 439 & 504 & 943 & 504 \\
\# internal edges & 527 & 734 & 1261 & 756 \\
\# external edges & 8 & 8 & -- & -- \\
$\langle k_\text{internal} \rangle$ & 2.40 & 2.91 & 2.69 & 3 \\
$\langle k_\text{external} \rangle$ & .0205 & .0179 & -- & -- \\
$\langle C \rangle$ & 0.0109 & 0.0821 & 0.0488 & 0.003(2) \\
$\langle \ell \rangle$ & 9.32 & 8.26 & 11.42 & 7.09(3)
  \end{tabular}
   \caption{Summary statistics of power grids $c$ and $d$ (in isolation and coupled together) and of a random 3-regular graph (``3-Regular'') with the same number of nodes as $d$. The statistics for the random 3-regular graph are averages over 1,000 realizations, with the standard deviation in parentheses to convey the fluctuation in the last digit.}
\end{table}

\subsection{Correlation between degree and load}
\begin{figure}[hbt]
\centerline{\includegraphics[width=7.23cm]{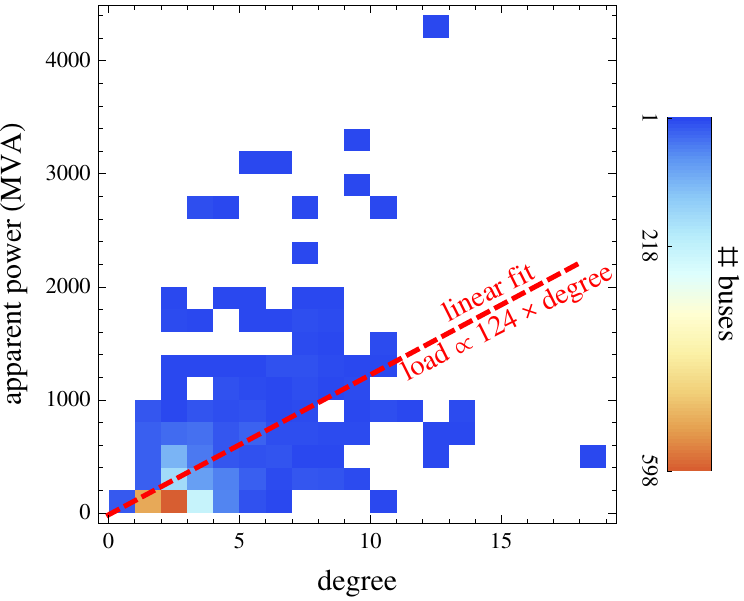}}
\label{c_d_degreeload}
\caption{Apparent power versus node degree for power grids $c$ and $d$, showing a weak correlation between the degree of nodes and their load. The best fit line $power =  -17.4+123.8 k$ indicates that each additional degree $k$ correlates with an increase in apparent power of 123.8 MVA ($R^2=0.30$).} 
\end{figure}

\ 
In the sandpile model studied here, we choose the capacity of every node to be its degree, so that toppled nodes unambiguously shed one grain to each neighbor. To examine the reasonableness of this assumption for real infrastructure, we sum the apparent power on the edges incident to each bus in power grids $c$ and $d$, and we plot in Fig.~\ref{c_d_degreeload} a density histogram of apparent power versus degree of the nodes. Most buses have low degree and low apparent power, and some buses have high degree but low power, while others bear high power among few connections. But the general trend is positive: each additional degree correlates with 123 MVA of additional apparent power, though the correlation is tenuous ($R^2=0.30$). This suggests that using the sandpile model with capacity equal to degree in order to study cascades of load in power grids---and by extension other infrastructure---is not unreasonable for the purpose of obtaining heuristic understanding.

\section{Branching process approximation}
\subsection{Comparing theory and experiment}\ 
We choose the dissipation of sand $f$ (the chance that a grain of sand is deleted as it is shed from one node to another) so that the largest cascades in simulations topple almost all the network. (As a rule of thumb, take $f = 20/N$, where $N$ is the total number of nodes.) Decreasing or increasing $f$ extends or shortens the tail of the avalanche size distribution, respectively. Thus we tune $f$ to utilize the entire system and to achieve a power law over as many scales as possible, since we see power law behavior in, e.g., blackouts in power grids~\cite{DobsonChaos2007}.

\begin{figure}[hbt]
\begin{center}
\centerline{\includegraphics[width=8.7cm]{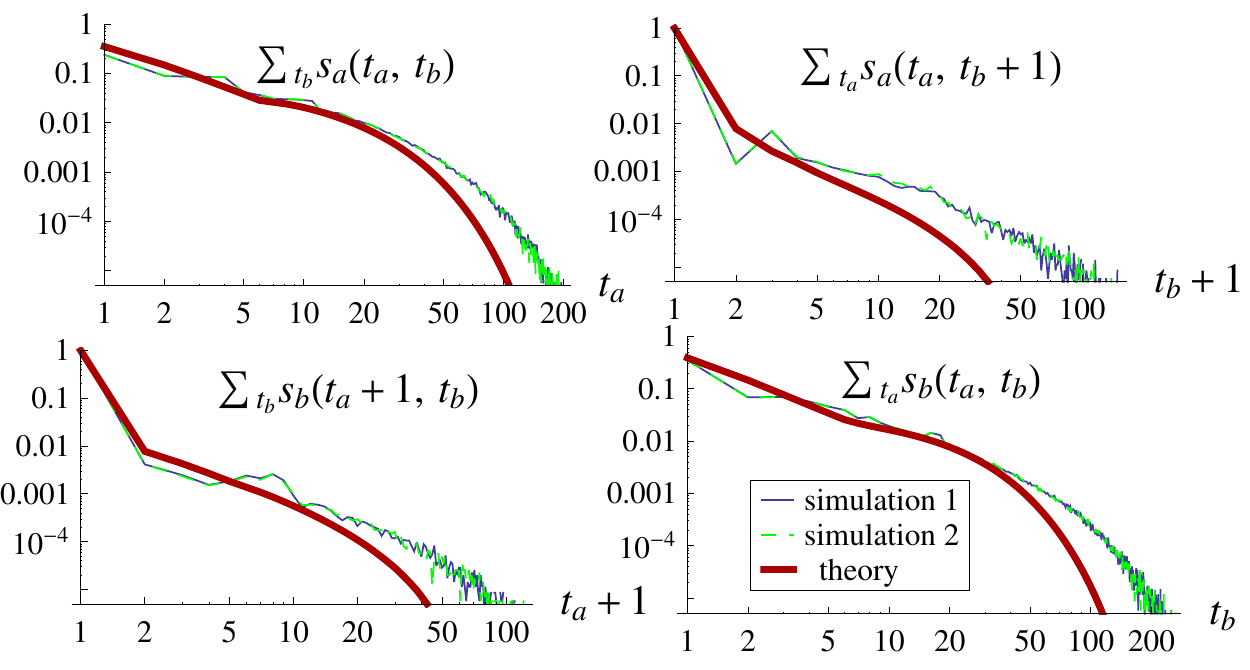}} 
\label{simulation_theory_cd_f05}
\caption{Blue and green curves: independent simulations of $10^6$ grains of sand dropped on power grids $c$ and $d$, labeled $a$ and $b$ respectively to match our theoretical notation. As in Fig.~3, we collect statistics after $10^5$ grains are dropped, initialized with amounts of sand chosen uniformly at random from zero to one less a node's degree, with dissipation of sand $f=0.05$ (half the value used in Fig.~3). Red curve: branching process approximation using the empirical degree distributions of $c$ and $d$, obtained by symbolically iterating the self-consistency equations (Eq.~\eqref{selfconsistencyequations}) seven times, expanding the result and collecting coefficients. The tails are decreased in the simulations due to finite system size (439 in $c$, 504 in $d$) and the dissipation of sand $f$, while the tails of the branching process approximation miss probability mass due of only iterating seven times (which took a week on a 2GHz laptop).}
\end{center}
\end{figure}

In Fig.~3 of the main paper we compare simulations of sandpile cascades with theoretical predictions for dissipation rate of sand $f=0.1$. This somewhat large dissipation rate mitigates the largest cascades. In Fig.~\ref{simulation_theory_cd_f05} we show results for the same branching process prediction (seven iterations of the self-consistency equations, started from $\mathcal{S}_a=\mathcal{S}_b=1$) with two independent simulations with dissipation rate of sand $f=0.05$. Halving the dissipation rate noticeably extends the tails of the marginalized avalanche size distributions---i.e., the largest avalanches become larger. Iterating the self-consistency equations an eighth time would 
more accurately compute the probability mass in the tails, but is beyond our computer resources. Showing the results of two independent simulations (blue and green curves of Fig.~\ref{simulation_theory_cd_f05}) illustrates the variation in the cascade size distributions. In particular, variation between simulations is apparent only in the tails of the distributions, as the largest avalanches are so rare.
Even though the behavior of sandpile cascades on single networks is rather robust to network structure~\cite{Goh_sandpile_PRL,Goh_sandpile_PhysicaA,Lee:2005fo,sandpile_ER,sandpile_wattsstrogatz_1d,sandpile_wattsstrogatz_2d}, we see sensitivity to the sparse connections among modules, as demonstrated in the main paper. What is more, the particular degrees of nodes with external links profoundly affect the inflicted avalanche size distributions, $\sum_{t_a} s_a(t_a,t_b)$ and $\sum_{t_b} s_b(t_a,t_b)$. Notice the characteristic ``blips'' (deviations from straight lines) that occur in the blue and green curves (the simulations) in the inflicted avalanche size distributions in the top-right and bottom-left plots in Fig.~\ref{simulation_theory_cd_f05}. For instance, there is a particularly high probability that an avalanche begun in $d$ topples 8, 9 or 10 nodes in $c$. We suspect this is due to the outlier node in network $c$ that has one external link and ten internal links---far more internal links than all other $c$-nodes with an external link. Consequently, a cascade in $d$ that leaks across to $c$ via this node topples with high probability around 8, 9 or 10 nodes in $c$ because of this node's high internal degree. Similarly, the blip at size 3 of the distribution of inflicted cascade size from $c$ to $d$ results from the average internal degree 2.5 of $d$-nodes with an external link. It is surprising, given the indifference of sandpile cascades to network structure, that the particular internal degrees of nodes with an external link greatly affect the marginalized avalanche size distributions.

\subsection{Distance between the avalanche size distributions and the product of their marginals}\ 
When we compare the branching process predictions with simulations, such as in Figs. 3 and \ref{simulation_theory_cd_f05}, we reduce the dimensions of the joint $s_a(t_a, t_b)$ by marginalizing in order to plot one-dimensional curves. This raises the question: how far are the products of the marginals, such as $(\sum_{t_a} s_a(t_a, t_b)) (\sum_{t_b} s_a(t_a, t_b))$, from the joint distributions? 
Using the first $11^2$ coefficients of $s_a(t_a, t_b)$ (for $0 \leq t_a, t_b \leq 11$) computed using Lagrange inversion (Theorem \ref{LagrangeTheorem}), we computed the distance between $s_a(t_a, t_b)$ and the product of its marginals---in the 2-norm, 1-norm and Kullback-Leibler divergence---as a function of Bernoulli coupling $p$ between two random 3-regular graphs. With increasing coupling $p$, the cascade sizes $t_a, t_b$ become increasingly correlated, and the joint distribution $s_a$ grows increasingly distant from the product of its marginals. This suggests that comparing marginals, as in Figs.~3 and~\ref{simulation_theory_cd_f05}, is an insufficient test of the branching process, and thus we next subject the branching process to its most stringent test.

\subsection{Fine-grained comparison of branching process and simulation}
\begin{figure}[hbt]
\begin{center}
\centerline{\includegraphics[width=8.7cm]{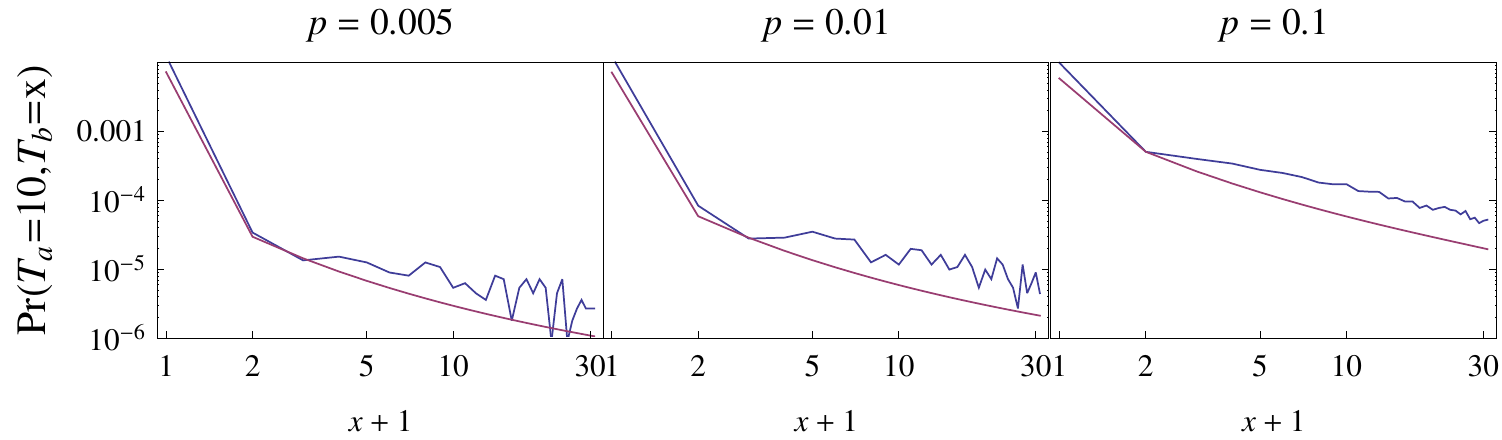}}
\label{comparesimtheorystrip}
\caption{Comparison of theory (red) and simulation (blue) of the chance of toppling 10 nodes in $a$ and $0 \leq x \leq 30$ nodes in $b$, in log-log scales. The simulations were Bernoulli-coupled random 3-regular graphs with $10^3$ nodes each, dissipation $f=0.02$, and $2\times 10^6$ grains dropped.}
\end{center}
\end{figure}

\ 
Since marginalizing $s_a, s_b$, as in Figs. 3 and \ref{simulation_theory_cd_f05}, may obscure deviation between theory and the branching process, we next compare the joint distribution $s_a(t_a, t_b)$ to simulation. 
In Fig.~\ref{comparesimtheorystrip} we plot $\Pr(T_a=10, T_b = x)$ for $0 \leq x \leq 30$ in the simulation (blue) and branching process prediction (red), for interconnectivity $p=0.005, 0.01, 0.1$. We compute the branching process prediction by calculating $s_a(t_a, t_b), s_b(t_a,t_b)$ for $t_a=10$ and $0 \leq t_b \leq 30$ using Lagrange inversion (Theorem \ref{LagrangeTheorem}). Since sand is dropped on nodes uniformly at random and each network has $10^3$ nodes, the avalanches begin in $a$ or in $b$ with equal probability, so that $Pr(T_a=10, T_b=x) = (s_a(10,x)+s_b(10,x))/2$. Although the branching process under-predicts theory for larger avalanches (especially with greater interconnectivity), we see qualitatively good agreement.

\subsection{Capacity disparity}\ 
Cascades inflicted from large capacity networks to small capacity networks are larger than those from small to large capacity networks. Here we derive the heuristic formula for this effect (Eq.~\eqref{inflictedratio}) using the multitype branching process, and we show qualitative agreement with simulation.

Differentiating the self-consistency equations \eqref{selfconsistencyequations} with respect to $\tau_a$ and $\tau_b$ and setting $\tau_a = \tau_b = 1$ yields four equations for the first moments of the avalanche size distributions $s_a, s_b$:
\begin{subequations} \label{s_moments}
\begin{align}
\langle s_a \rangle_a &= \frac{\left\langle u_b\right\rangle _b-1}{\gamma}, \quad \langle s_b \rangle_b = \frac{\left\langle u_a\right\rangle _a-1}{\gamma}, \\
\langle s_a \rangle_b &= \frac{-\left\langle u_a\right\rangle _b}{\gamma}, \quad \langle s_b \rangle_a = \frac{-\left\langle u_b\right\rangle _a}{\gamma}, \label{inflicted}
\end{align}
\end{subequations}
with common denominator
\begin{align} \label{denominator}
\gamma = -1+\left\langle u_a\right\rangle _b \left\langle u_b\right\rangle _a-\left\langle u_a\right\rangle _a \left(\left\langle u_b\right\rangle _b -1 \right)+\left\langle u_b\right\rangle _b.
\end{align}
For the first moments of the branch distributions $u_a, u_b$ for \RBR{z_a}{p}{z_b},
\begin{align}\label{RBR_u_moments}
\left\langle u_a\right\rangle _a &=1-\frac{p}{1+z_a},\quad \left\langle u_a\right\rangle _b=\frac{p}{1+z_b}
\end{align}
(and vice versa for $u_b$, interchanging $a$ with $b$), the denominator \eqref{denominator} is zero, and Eqs. \eqref{s_moments} are all infinite. This is not a surprise, because in isolation the networks' cascade size distributions are asymptotically power laws with exponent $-3/2$, which have infinite first moments. However, one can compare the rates of divergence of the mean inflicted cascade sizes, Eqs. \eqref{inflicted}, by computing their ratio, $\langle s_a \rangle_b  / \langle s_b \rangle_a = (1+z_a)/(1+z_b)$. No other ratios appear to contain useful information (nor dependence on $p$ with which to estimate the critical amount of interconnectivity $p^*$):
\begin{align*}
\frac{\langle s_a \rangle_a}{\langle s_b \rangle_a} = \frac{\langle s_a \rangle_a}{\langle s_b \rangle_b} = \frac{1+z_a}{1+z_b}, \quad \frac{\langle s_a \rangle_a}{\langle s_a \rangle_b} = 1.
\end{align*}

\begin{figure}[htbp]
\centerline{\includegraphics[width=8.7cm]{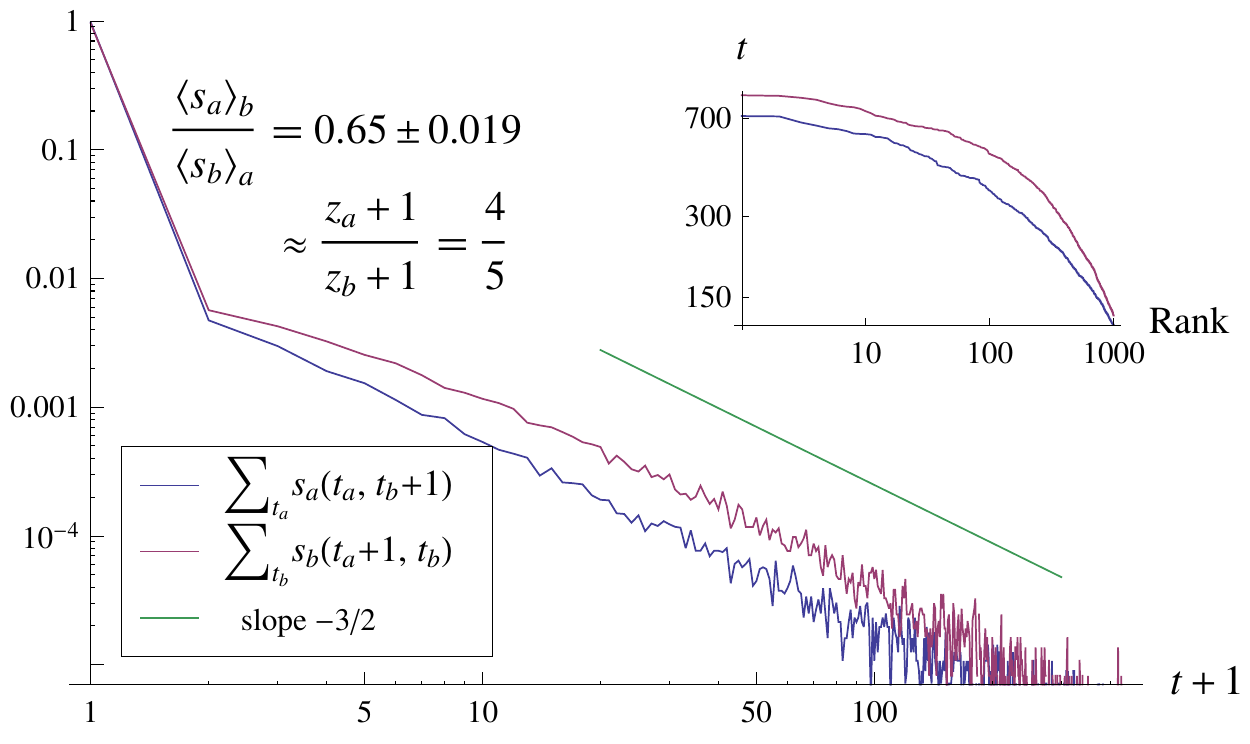}}
\label{asymmetriccapacity}
\caption{Qualitative agreement between the theoretical prediction (Eq.~\eqref{inflictedratio}) and simulation on \RBR{3}{0.005}{4} ($2\times10^6$ grains, $10^3$ nodes/network, $f=0.02$). Main plot: marginalized avalanche size distributions of inflicted cascades from $a$ to $b$ (blue), $b$ to $a$ (red). Inset: rank-size plot of the largest $10^3$ inflicted cascades from $a$ to $b$ (blue), $b$ to $a$ (red).}
\end{figure}

In all simulations on \RBR{z_a}{p}{z_b} with $z_a \neq z_b$, the heuristic formula Eq.~\eqref{inflictedratio} works qualitatively, in that inflicted cascades are always larger from the larger capacity network (i.e., $z_a < z_b \implies \langle s_{a}\rangle_b < \langle s_{b} \rangle_a$). But finding quantitative agreement with Eq.~\eqref{inflictedratio} is difficult because the avalanche size distributions are power laws with such heavy tails that their variances diverge. Figure~\ref{asymmetriccapacity} is an example: for \RBR{3}{0.005}{3}, $\langle s_{a}\rangle_b = 0.086 \pm 0.019$, and $\langle s_{b}\rangle_a = 1.32 \pm 0.023$ (mean $\pm$ standard error), which yields $\langle s_{a}\rangle_b / \langle s_{b} \rangle_a = 0.65 \pm 0.019$, while theory predicts $(1+z_a)/(1+z_b) = 4/5$ (20\% error). Nevertheless, Eq.~\eqref{inflictedratio} provides a simple, useful heuristic for the dangers of small capacity networks connecting to large capacity networks.

\subsection{Computational challenges in showing the locally stabilizing effect}
\begin{figure}[hbt]
\begin{center}
\centerline{\includegraphics[width=8.7cm]{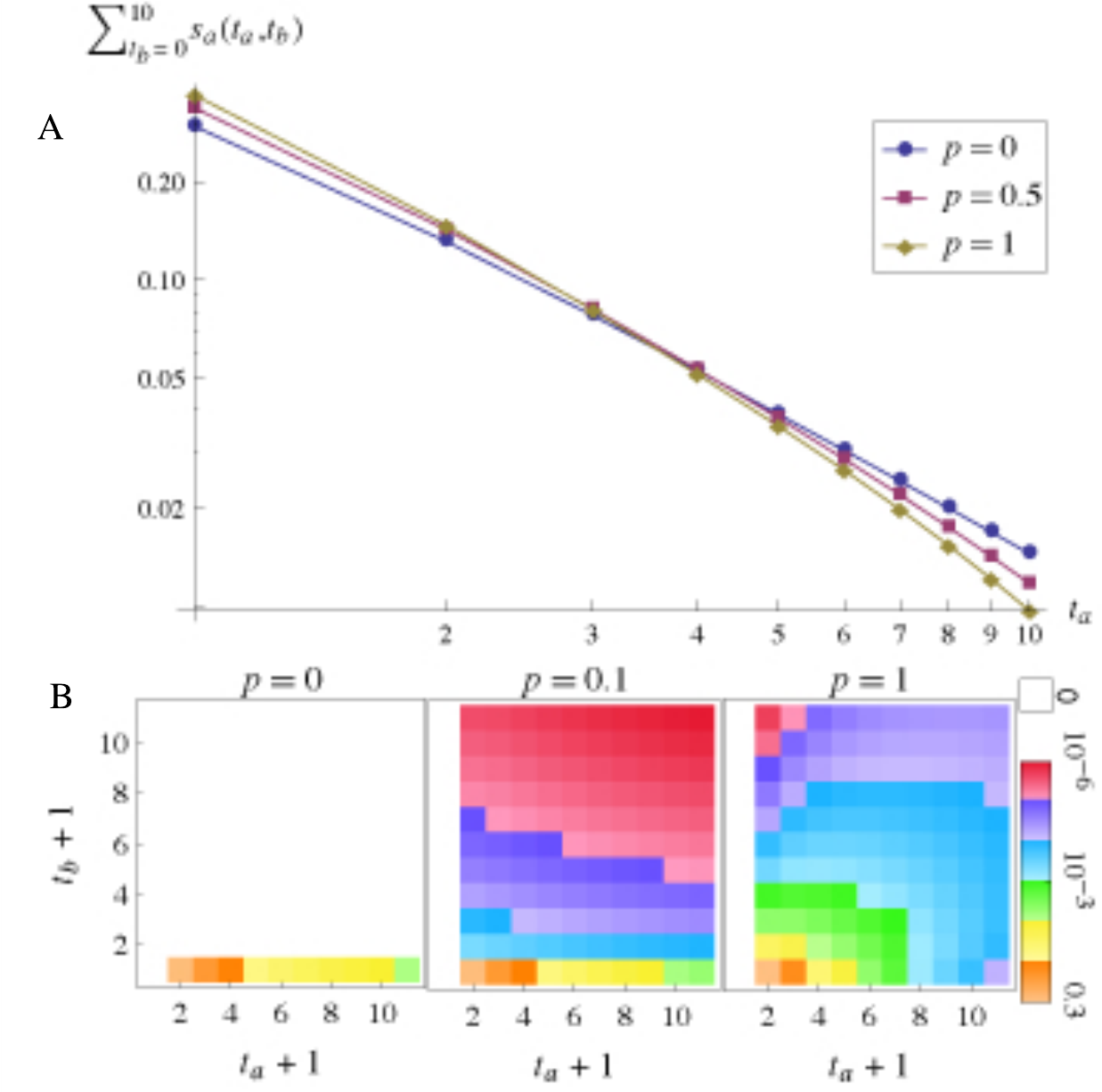}}
\label{IJGoodfigures}
\caption{Panel A: Increasing the interconnectivity $p$ between two random 3-regular graphs appears to mitigate the tails of the local avalanche size distribution $\sum_{t_b} s_a(t_a,t_b)$. Plotted are the first $11 \times 11$ coefficients of $\mathcal{S}_a$ computed using multidimensional Lagrange inversion \cite{Good_Lagrange} for $p=0,0.5,1$. Panel B: Increasing the interconnectivity $p$ between two random 3-regular graphs ``smears out'' the avalanche size distribution $s_a(t_a, t_b)$, so that cascades large in both networks become more likely. Plotted are $s_a(t_a, t_b)$ for $0 \leq t_a, t_b \leq 10$, computed using multidimensional Lagrange inversion \cite{Good_Lagrange} for $p=0,0.1,1$. We plot zero probabilities in white and positive probabilities in a logarithmic color scale from orange (high probability) to bright red (low probability).}
\end{center}
\end{figure}

\ Although the locally stabilizing effect of interconnections is apparent in simulations, demonstrating it analytically poses mathematical and computational challenges. Here we use multidimensional Lagrange inversion (Theorem \ref{LagrangeTheorem} in the {\it Materials and Methods}) to solve for the probabilities of the smallest avalanches, $s_a(t_a, t_b)$ for $0 \leq t_a, t_b \leq 10$. Next we plot in Fig.~\ref{IJGoodfigures}A the marginalized avalanche size distribution $\sum_{t_b = 0}^{10} s_a(t_a,t_b)$.

Although it appears in Fig.~\ref{IJGoodfigures}A that the largest avalanches become less likely with increasing interconnectivity $p$, in accordance with the results in simulations, this figure has a caveat: we have only computed up to avalanche size at most 10 in each network. As a result, the right-hand tail of Fig.~\ref{IJGoodfigures}A lacks probability mass compared to the actual marginalized avalanche size distribution $\sum_{t_b = 0}^{\infty} s_a(t_a,t_b)$, because, for example, there is a significant chance of an avalanche of size $t_a=10$ and $t_b=11, 12$ or $13$. This figure could be improved by computing  $s_a(t_a, t_b)$ along ``vertical strips'' $0 \leq t_b \leq t_b^{max}$, where $t_b^{max} \gg 1$, for just a handful of values $t_a$. However, computing the large coefficients of $\mathcal{S}_a$ requires differentiating expressions raised to large powers, so the largest coefficients are the most difficult to compute.

The Cauchy formula, Eq.~\eqref{Cauchyintegral} of the {\it Materials and Methods}, uses integration rather than differentiation, which can be more numerically stable \cite{watts_newman_strogatz}. But computing the largest coefficients $s_a(t_a, t_b)$ with this method requires integrating an increasingly large expression, namely the result of iterating the self-consistency equations at least $t_a+t_b+1$ times, starting from $\mathcal{S}_a=\mathcal{S}_b=1$. This, too, takes a long amount of computation time: computing $s_a(t_a, t_b)$ for $0 \leq t_a, t_b \leq 20$ on a typical laptop computer would take on the order of a week.

Nonetheless, even the smallest $11^2$ coefficients of $s_a$ and $s_b$ are useful. Fig. \ref{IJGoodfigures}B, for example, shows the joint avalanche size distribution $s_a(t_a, t_b)$ for two random regular graphs with Bernoulli($p$)-distributed coupling, for $p=0,0.1, 1$. As the interconnectivity $p$ increases, the cascades begun in $a$ increasingly ``smear out'' among $a$ and $b$, and cascades are more frequently large in both networks at once.



\end{document}